\documentclass[12pt,english]{article}
\usepackage[LGR,T1]{fontenc}
\usepackage{textcomp}
\usepackage[utf8]{inputenc}
\usepackage{amsmath}
\usepackage{cancel}
\usepackage{stackrel}
\usepackage{graphicx}
\usepackage{geometry}
\usepackage{setspace}
\usepackage{slashed}
\usepackage{bbm}
\allowdisplaybreaks[4]

\providecommand{\tabularnewline}{\\}

\usepackage{babel}

\author{
Asif Ali$^\text{a}$,
Yi-Jie Li$^\text{a}$,
Guang-Zhi Xu$^\text{a,1}$,
Kui-Yong Liu$^\text{b,a,1}$
\\[8pt] 
\parbox{\linewidth}{\small
$^\text{a}$ School of Physics, Liaoning University, Shenyang 110036, China\\
$^\text{b}$ School of Physics and Electronic Technology, Liaoning Normal University, Dalian 116029, China\\
$^\text{1}$ Corresponding authors: \texttt{xuguangzhi@lnu.edu.cn}, \texttt{liukuiyong@lnu.edu.cn}
}
}

\makeatother

\usepackage{babel}
\begin{document}
\title{Rare Radiative Decays of Z boson into S- and P-wave Charmonium within
the Bethe-Salpeter Model}
\maketitle
\begin{abstract}
We employ the Bethe–Salpeter (BS) model to investigate radiative decays 
of the Z boson into charmonium. Within the instantaneous 
approximation of the BS equation, we analyze the relevant decay channels for S-wave 
charmonium states ($\psi(nS)$, $\eta_{c}(nS)$, $n=1,2,3$) and P-wave charmonium states 
($\chi_{c0}(mP)$, $\chi_{c1}(mP)$, $h_{c}(mP)$, $m=1,2$). 
We compute all branching fractions using leading-order Feynman diagrams. 
For the $J/\psi$ channel, our BS calculations suggest that prior theoretical results were likely underestimated, 
making our prediction more promising for future experimental measurements.
\end{abstract}

\section{Introduction}

Quarkonium production and decay processes are of theoretical and experimental
interest. They provide remarkable opportunities to understand quantum
chromodynamics (QCD) and investigate the dynamics and structure of
quarkonium. Understanding the interplay between short-distance and
long-distance effects in QCD remains a fundamental challenge in particle
physics. At short distance, asymptotic freedom ensures the strong
coupling $\alpha_{s}$ is small, allowing for reliable perturbative
calculations. In the long-distance regime, however, where $\alpha_{s}\sim1$,
non-perturbative approaches are required.

In studies of heavy quarkonium production and decay, one widely used
approach is the non-relativistic QCD (NRQCD) factorization formalism \cite{Bodwin1997}. 
This model relies on the fact
that the mass of a quark, $m$, is significantly larger than the typical
scale of the strong interaction $\Lambda_{QCD}$. 
Consequently, the relative internal velocity of the quark--antiquark pair is small enough 
to permit a systematic expansion of production ratios and decay widths in powers of the relative quarkonium velocity.
At leading-order (LO) in this expansion,
the internal motion of the quarks is neglected entirely. Moreover,
the long-distance part of the interaction is then described by coefficients
that can be determined, for instance, through potential models or
by fitting experimental data. However, it is important to note that
for charmonium, the expansion parameter is not particularly small
($v\sim0.3$), so the LO approximation may introduce significant uncertainties.
A notable example is the production of double charmonium in $e^{+}e^{-}$
annihilation at B-factories \cite{BaBar:2005nic,Belle:2004abn}, where the
LO NRQCD predictions \cite{Liu:2002wq} were found to be approximately
an order of magnitude smaller than the experimental measurements.
Therefore, reliable predictions require including NLO radiative and
relativistic corrections to the LO contribution. 

An alternative framework to investigate charmonium production and
decay is provided by the Bethe-Salpeter (BS) formalism \cite{Salpeter1952},
which is a non-perturbative approach for studying relativistic bound
state problems. Firmly rooted in quantum field theory, and grounded
in dynamical equations, it offers a realistic description for understanding
quarkonium as composite objects, and a useful tool to study low- and
high-energy production processes involving quarkonia. It is an elegant
model that comprehensively describes the relativistic effects of quark
spins as well as internal quark motion within the meson in a relativistically
covariant way. Heavy quarkonium production and decay have been studied
using the BS model, yielding reliable results \cite{Geng2019,Li2024,Negash2019}.

Experimental searches for radiative Z-boson decays into a single heavy
quarkonium have been conducted by the ATLAS and CMS collaborations
at the LHC \cite{ATLAS:2022rej,ATLAS:2015vss}; however, no significant signal
has been observed to date. In recent years, several high-luminosity
$e^{+}e^{-}$ colliders, such as the ILC \cite{ILC:2013jhg}, FCC-ee
\cite{FCC:2018evy}, Super Z factory \cite{SuperZfactory2010}, and CEPC \cite{CSG2018},
have been planned to operate at the Z pole. This will provide new
opportunities to study rare Z boson decays with unprecedented precision.
Moreover, several theoretical approaches have been developed to examine
these processes. In Ref. \cite{Luchinsky2017}, the authors employed
both the NRQCD and light-cone distribution amplitude (LCDA) approaches
to investigate the radiative decays of the Z boson into S-wave or
P-wave charmonium. Similarly, the decays $Z\rightarrow J/\psi(\Upsilon)+\gamma$
were studied in both the NRQCD and LCDA approaches in Ref. \cite{Huang2015},
and they also evaluated the indirect contribution where the Z-boson
decays into a real photon and a virtual photon through a heavy-quark
or W-boson loop, with the virtual photon fragmenting into a heavy
meson. Recently, the Z-boson decays into a photon plus S-wave quarkonium
have been analyzed up to $O(\alpha_{s}v^{2})$ corrections \cite{Wang:2023ssg}.
Moreover, the decay rates for Z boson into photon plus S-wave or P-wave
quarkonium have been studied up to next-to-next-to-leading order(NNLO)
in $\alpha_{s}$ within the NRQCD framework \cite{Sang2022,Sang2023}.

For the first time, we employ the full instantaneous BS formalism to systematically investigate both S-wave and P-wave charmonium final states in Z radiative decays, 
extending our earlier heavy-quark limit study of S-wave channels \cite{Ali:2026ria}.
In the present work, our aim is to obtain reliable predictions by incorporating both relativistic and perturbative corrections. Unlike simple potential models,
the BS equation enables a consistent treatment of quark–antiquark bound states.
Our analysis therefore helps to elucidate the impact of relativistic corrections on 
these rare yet phenomenologically significant decay processes.

Our paper is organized as follows. In Sec.~\ref{sec:bsf}, we introduce the BS framework and a proper quark-antiquark QCD-inspired kernel. In Sec.~\ref{sec:decaywidths}, we derive the BS amplitudes and decay widths for the radiative decays of the Z boson into quarkonia, including S-wave states ($J^{PC}=0^{-+}, 1^{--}$) and P-wave states ($J^{PC}=J^{++}$ with $J=0,1$ and $J^{PC}=1^{+-}$). In Sec.~\ref{sec:numres}, we present the numerical branching-fraction results and compare them with previous works. Finally, we summarize our main conclusions in Sec.~\ref{sec:sum}.

\section{Bethe-Salpeter Framework }
\label{sec:bsf}

We begin with the basic four-dimensional (4D) BS equation for a $Q\bar{Q}$ bound
system. 
\begin{equation}
S^{-1}(p_{1})\Psi(q)S^{-1}(-p_{2})=i\int\frac{\mathrm{d}^{4}k}{(2\pi)^{4}}K(q,k)\Psi(k)\label{eq:bs4d}
\end{equation}

Here, $\Psi(q)$ denotes the 4D BS wavefunction and $K(q,k)$ is the
internal kernel, which defines the interaction between quark and anti-quark,
while $S^{-1}(\pm p_{1,2})$ are inverse propagators of quark and
antiquark with momenta denoted by $p_{1}$ and $p_{2}$ respectively.
$q$ is the internal relative momentum of the quark and antiquark
in the meson with mass $M$ and external momentum $P$. For the equal-mass
system, quark and antiquark momenta are related to the external and
internal momenta via 
\[
p_{1,2\mu}=\frac{1}{2}P_{\mu}\pm q_{\mu}.
\]

Under the instantaneous approximation, the BS kernel becomes 
\[
K(q,k)=K(q_{\perp},k_{\perp}).
\]
where the transverse component $q_{\perp\mu}$ is written as $q_{\perp\mu}=q_{\mu}-\frac{q\cdot P}{M^{2}}P_{\mu}$,
orthogonal to the meson momentum, i.e. $q_{\perp}\cdot P=0$. The
longitudinal component to the $P_{\mu}$ is $q_{\parallel}=\frac{q\cdot P}{M^{2}}P_{\mu}$.
Accordingly, the internal relative momentum can be decomposed as $q_{\mu}=(q_{\parallel},q_{\perp})$
where $q_{\parallel}$ corresponds to the temporal component. Correspondingly,
we define two Lorentz invariant variables: 
\begin{align*}
q_{P}=P\cdot q/M & ,q_{T}=\sqrt{q_{P}^{2}-q^{2}}=\sqrt{-q_{\perp}^{2}}.
\end{align*}
In the center-of-mass frame, $\vec{P}=0$, and these quantities reduce
to the usual component $q^{0}$ and $|\vec{q}|$, respectively. 
We subsequently integrate over the longitudinal component of the 4D volume element
$\mathrm{d}^{4}k$ appearing on the right side of Eq.~\ref{eq:bs4d}, yielding 
\begin{equation}
S^{-1}(p_{1})\Psi(q)S^{-1}(-p_{2})=\int\frac{\mathrm{d}^{3}k_{\perp}}{(2\pi)^{3}}K(q_{\perp},k_{\perp})\psi(k_{\perp})=\Gamma(q_{\perp}).\label{eq:c}
\end{equation}

Here, we define 
\begin{equation}
\psi(k_{\perp})=\frac{i}{2\pi}\int\mathrm{d}k_{P}\Psi(k),\label{eq:b}
\end{equation}

and $\Gamma(q_{\perp})$ represents the hadron-quark vertex function.
Finally, the BS wavefunction can then be expressed as 
\begin{equation}
\Psi(q)=S_{1}(p_{1})\Gamma(q_{\perp})S_{2}(-p_{2})\label{eq:d}
\end{equation}

The fermion propagator of the quarks takes the decomposed form as
\begin{align}
S_{i}(\pm p_{i}) & =\frac{\varLambda_{i}^{+}(q_{\perp})}{j(i)q_{P}+\frac{1}{2}M-\omega_{j}}+\frac{\varLambda_{i}^{-}(q_{\perp})}{j(i)q_{P}+\frac{1}{2}M+\omega_{j}}.\label{popd}
\end{align}

In the above expression, 
\begin{align}
\omega_{j}^{2}=m_{i}^{2}-q_{\perp}^{2}=m_{i}^{2}+\vec{q}^{2},\text{and }\varLambda_{i}^{\pm} & =\frac{1}{2\omega_{i}}[\frac{\slashed{P}}{M}\omega_{i}\pm j(i)(m_{i}+\slashed{q}_{\perp})].\label{eq:n1}
\end{align}
Here $i=1,2$ corresponds to the quark and antiquark, respectively,
with $j(i)=(-1)^{i+1}$. In Eq.~\ref{eq:n1} $\varLambda_{i}^{\pm}$
acts on $\psi(q_{\perp}),$ giving the projected wave functions. 
\begin{align}
\psi^{\pm\pm}(q_{\perp}) & =\varLambda_{1}^{\pm}(q_{\perp})\frac{\slashed{P}}{M}\psi(q_{\perp})\frac{\slashed{P}}{M}\varLambda_{2}^{\mp}(q_{\perp}).\label{projwave}
\end{align}

After performing contour integration over $q_{P}$ in Eq.~\ref{eq:d},
and simplifying the resulting expressions, we obtain four independent
Salpeter equations. 
\begin{align}
(M-\omega_{1}-\omega_{2})\psi^{++}(q_{\perp}) & =\varLambda_{1}^{+}(q_{\perp})\Gamma(q_{\perp})\varLambda_{2}^{+}(q_{\perp})\nonumber \\
(M+\omega_{1}+\omega_{2})\psi^{--}(q_{\perp}) & =-\varLambda_{1}^{-}(q_{\perp})\Gamma(q_{\perp})\varLambda_{2}^{-}(q_{\perp})\label{eq:4inde}\\
\psi^{+-}(q_{\perp}) & =0\nonumber \\
\psi^{-+}(q_{\perp}) & =0.\nonumber 
\end{align}

The corresponding normalization condition reads 
\begin{equation}
\int\frac{\mathrm{d}^{3}\vec{q}}{(2\pi)^{3}}Tr\left[\bar{\psi}^{++}(q_{\perp})\frac{\slashed{P}}{M}\psi^{++}(q_{\perp})\frac{\slashed{P}}{M}-\bar{\psi}^{--}(q_{\perp})\frac{\slashed{P}}{M}\psi^{--}(q_{\perp})\frac{\slashed{P}}{M}\right]=2P_{0}.\label{eq:normf-1}
\end{equation}
To solve the Salpeter equations, we adopt a composite interaction
kernel containing two components, the linear scalar confinement potential
and vector potential originating from single-gluon exchange. The momentum-space
expression of the kernel is written as \cite{Chang2010} 
\begin{align}
V(\vec{q})= & V_{v}(\vec{q})+\gamma_{0}\otimes\gamma^{0}V_{s}(\vec{q})\nonumber \\
V_{v}(\vec{q})= & \frac{2\pi\alpha_{s}(\vec{q}^{2})}{3\pi^{2}(\vec{q}^{2}+\beta^{2})}\nonumber \\
V_{s}(\vec{q})= & -(\frac{\lambda}{\beta}+V_{0})\delta^{3}(\vec{q})+\frac{\lambda}{\pi^{2}}\frac{1}{(\vec{q}^{2}+\beta^{2})^{2}}\label{eq:ker}
\end{align}
where the one-loop QCD running coupling constant is given by 
\[
\alpha_{s}(\vec{q}^{2})=\frac{12\pi}{(33-2N_{f})}\frac{1}{\ln(\mathbbm{e}+\vec{q}^{2}/\varLambda_{QCD}^{2})}.
\]
with $N_f=3$ as active quark flavors at the charmonium scale.
The constants $\lambda$,
$\beta$, $\mathbbm{e}$, $\Lambda_{QCD},$ and $V_{0}$ are adjustable quantities
that determine the profile of the interaction potential. Among them,
$V_{0}$ is determined by fitting the mass of the lowest radial state
of the corresponding quarkonium states.

\section{Derivation of Radiative Decay Widths for Z Boson Decays to Charmonium}
\label{sec:decaywidths}

This section is devoted to investigating the formulas of decay widths
for the radiative decays of the Z boson into Charmonium. We have constructed
the most general form of the relativistic BS wave functions with definite
parity, charge conjugation, and total angular momentum for these charmonium
states. The corresponding Salpeter equations have been solved using
an appropriate QCD-inspired quark--antiquark kernel for charmonium.
This yields the radial BS wave functions in momentum space, which
are then used to compute the decay widths for the respective quarkonium
states.

\begin{figure}[h]
\centering \includegraphics[scale=0.8]{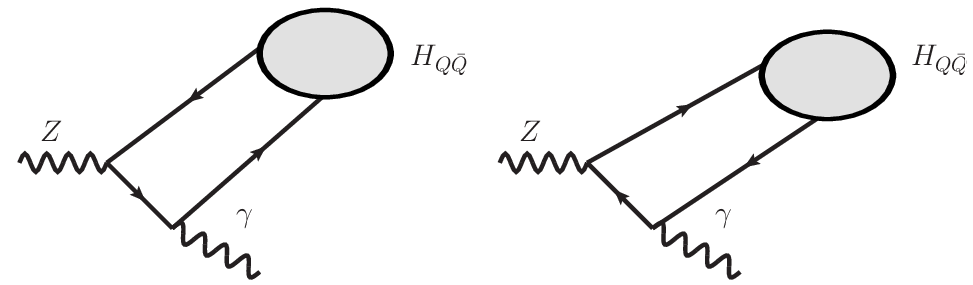} 
\caption{Leading-order Feynman diagrams for the $Z\rightarrow H_{Q\bar{Q}}\gamma$
process, where $H_{Q\bar{Q}}$ denotes a charmonium state. }
\label{fig:feyndia} 
\end{figure}

The LO Feynman diagrams of the $\alpha_{s}$ expansion, which describe purely electroweak processes, are presented in Fig.~\ref{fig:feyndia}.
In these diagrams, a $Z$ boson produces a photon with momentum
$K$, and a quarkonium state with momentum $P$. 
The vertex factor for the $ZQ\bar{Q}$ interaction reads
$-i\frac{g}{2\cos\theta_{W}}\gamma^{\mu}\big(g_{v}^{Q}-g_{a}^{Q}\gamma^{5}\big)$,
where $T_{3}^{Q}$ labels the third component of the heavy quark’s weak isospin, such that $g_{a}^{Q}=T_{3}^{Q}$ and $g_{v}^{Q}=T_{3}^{Q}-2e_{Q}\sin^{2}\theta_{W}$.
The $\mathrm{SU}(2)_{L}$ gauge coupling $g$ is related to the electromagnetic coupling $e$ via $g = e/\sin\theta_{W}$, with $e^{2}=4\pi\alpha_{\mathrm{EM}}$.
Here $\alpha_{\mathrm{EM}}$ is the electromagnetic fine-structure constant, and $\theta_{W}$ denotes the Weinberg mixing angle.
We include the individual contributions from both diagrams together with the interference terms between them.

\subsection{$0^{-+}$ Charmonium}

We have studied the decay channels of the pseudoscalar S-wave charmonium,
namely $Z\rightarrow\eta_{c}(nS)+\gamma$, within the BS formalism,
with $n=1,2,3$. Due to charge conjugation invariance, the production
of $\eta_{c}+\gamma$ proceeds via the vector coupling. The corresponding
4D-BS amplitude is given by the momentum integral as 
\begin{align}
\mathcal{M}_{Z\rightarrow\eta_{c}\gamma} & =i\frac{\sqrt{N_{c}}ee_{Q}gg_{v}^{Q}}{2\cos\theta_{W}}\text{\ensuremath{\int\frac{\mathrm{d}^{4}q}{(2\pi)^{4}}}Tr}
[\bar{\Psi}_{\eta}(q)\{\slashed{\epsilon}_{\gamma}\frac{1}{\slashed{K}+\slashed{p}_{1}-(m-i\epsilon)}\slashed{\varepsilon}_{Z}+\slashed{\varepsilon}_{Z}\frac{1}{-\slashed{K}-\slashed{p}_{2}-(m-i\epsilon)}\slashed{\epsilon}_{\gamma}\}]\label{eq:am1}\\
 & =\mathcal{M}_{1}+\mathcal{M}_{2}\nonumber 
\end{align}
In the above expression, $N_{c}$ is the number of colors of the QCD
$\mathrm{SU}(3)$ color group. $m$ is the mass of charm quark. $\Psi_{\eta}(q)$
is the relativistic 4D-BS wavefunction of the pseudoscalar meson,
and its corresponding conjugate is defined as $\bar{\Psi}\equiv\gamma^{0}\Psi^{\dagger}\gamma^{0}$.
Here, $\mathcal{M}_{1}$ and $\mathcal{M}_{2}$ are the amplitudes
corresponding to the left- and right-hand side diagrams of Fig.~\ref{fig:feyndia},
respectively.

We begin with the general form of BS wave function for the $\eta_{c}$.
It can be expressed in terms of various Dirac structures \cite{Chang2010,Bhatnagar:2024ykb,Kim:2003ny}.
The general decomposition form of the instantaneous BS wave function
for pseudoscalar meson ($J^{PC}=0^{-+}$ ) is 
\begin{equation}
\psi_{\eta}(q_{\perp})=[\slashed{P}\phi_{1}(q_{\perp})+M\phi_{2}(q_{\perp})+\frac{\slashed{P}\slashed{q}}{M}\phi_{3}(q_{\perp})]\gamma^{5}.\label{eq:PSED-1}
\end{equation}
In the above equation $\phi_{1},\phi_{2}$ are LO terms, whereas the
$\phi_{3}$ term corresponds to relativistic corrections. Moreover,
since $\phi_{1}\sim\phi_{2}$ numerically, setting $q\sim0$ in Eq.~\ref{eq:PSED-1},
reduces it to the Schrödinger wavefunction. Thus, $q$ characterizes
the magnitude of the relativistic effects in the quarkonium system.
By substituting Eq.~\ref{eq:PSED-1} into the last two equations
of Eq.~\ref{eq:4inde} and taking Dirac traces on both sides of the
resulting equations, we obtain 
\begin{equation}
\phi_{3}(q_{\perp})=-\frac{M}{m}\phi_{1}(q_{\perp}).
\end{equation}

\begin{figure}[h]
\centering \includegraphics[scale=0.4]{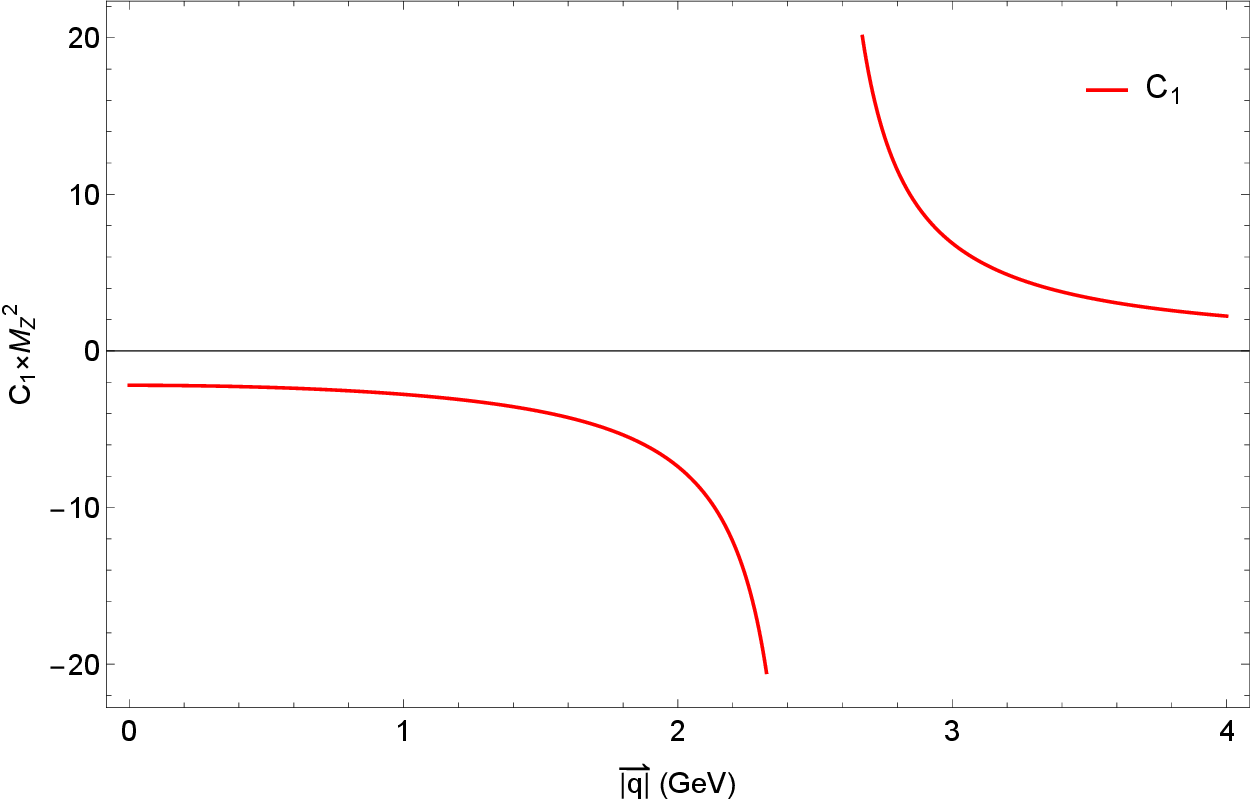}\includegraphics[scale=0.4]{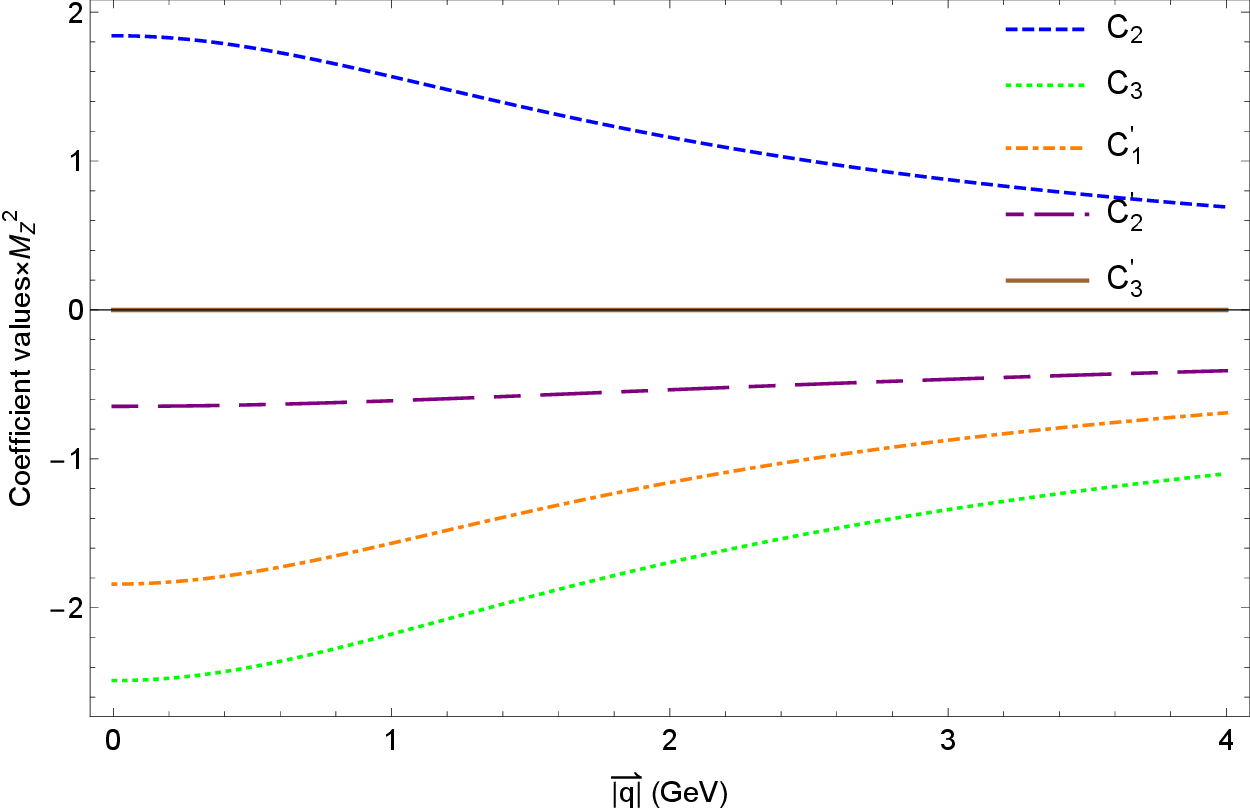}
\caption{ Coefficients $C_{1}$ (left) and $C_{i},\,C_{i}'$ (right) multiplied
by $M_{Z}^{2}$ versus $|\vec{q}|$ in the range $0<|\vec{q}|<4\,\text{GeV}$
for the $\eta_{c}$ state.
\label{fig:ceta}
}
\end{figure}

As a result, the wavefunction turns into 
\begin{equation}
\psi_{\eta}(q_{\perp})=[\slashed{P}\phi_{1}(q_{\perp})+M\phi_{2}(q_{\perp})-\frac{\slashed{P}\slashed{q}}{m}\phi_{1}(q_{\perp})]\gamma^{5}.\label{etawf}
\end{equation}
And the corresponding normalization condition is given by 
\begin{equation}
4M^{2}\int\frac{\mathrm{d}^{3}\vec{q}}{(2\pi)^{3}}\frac{\phi_{1}\phi_{2}\omega}{m}=P^{0}.
\end{equation}
By substituting Eq.~\ref{eq:PSED-1} into the first two equations
of Eq.~\ref{eq:4inde} and taking Dirac traces on both sides of the
resulting equations, we obtain the independent coupled integral equations
\begin{align}
(M-2\omega)\Bigl[\phi_{1}(q_{\perp})+\phi_{2}(q_{\perp})\frac{m}{\omega}\Bigr]= & -\int\frac{\mathrm{d}^{3}\vec{k}}{(2\pi)^{3}\omega^{2}}\nonumber \\
 & \left[(V_{s}-V_{v})(\phi_{1}(k_{\perp})m^{2}+\phi_{2}(k_{\perp})m\omega)-(V_{s}+V_{v})\phi_{1}(k_{\perp})\vec{k}\cdot\vec{q}\right]\\
(M+2\omega)\Bigl[\phi_{1}(q_{\perp})-f_{2}(q_{\perp})\frac{m}{\omega}\Bigr]= & \int\frac{\mathrm{d}^{3}\vec{k}}{(2\pi)^{3}\omega^{2}}\nonumber \\
 & \left[(V_{s}-V_{v})(\phi_{1}(k_{\perp})m^{2}-\phi_{2}(k_{\perp})m\omega)-(V_{s}+V_{v})\phi_{1}(k_{\perp})\vec{k}\cdot\vec{q}\right]
\end{align}
We can solve the above equations numerically for a given kernel to
obtain the $\phi_{1}$ and $\phi_{2}$, as well as mass spectra, as
an eigenvalue problem. Now, we aim to reduce each amplitude in Eq.~\ref{eq:am1}
to 3D form. Let's begin with 
\begin{align}
\mathcal{M}_{1} & =i\frac{\sqrt{3}ee_{Q}gg_{v}^{Q}}{2\cos\theta_{W}}\text{\ensuremath{\int\frac{\mathrm{d}^{4}q}{(2\pi)^{4}}}Tr}[\bar{\Psi}_{\eta}(q)\slashed{\epsilon}_{\gamma}\frac{1}{\slashed{K}+\slashed{p}_{1}-(m-i\epsilon)}\slashed{\varepsilon}_{Z}]\nonumber \\
 & =i\frac{\sqrt{3}ee_{Q}gg_{v}^{Q}}{2\cos\theta_{W}}\text{\ensuremath{\int\frac{\mathrm{d}^{4}q}{(2\pi)^{4}}}Tr}[\bar{\Psi}_{\eta}(q)\slashed{\epsilon}_{\gamma}\frac{\big((1/2+q_{P}/M)\slashed{P}+\slashed{q}_{\perp}+\slashed{K}+m\big)}{(q_{P}-b_{1}-i\epsilon)(q_{P}-b_{2}+i\epsilon)}\slashed{\varepsilon}_{Z}]\label{M14d}
\end{align}
The relative momentum $q^{\mu}$ can be decomposed into components
perpendicular and parallel to $P_{1}^{\mu}$ as $q_{\mu}=q_{\perp}{}_{\mu}+\frac{q_{P}}{M}P_{\mu}$
. Moreover, in the center-of-mass frame where $\eta_{c}$ and photon
are emitted back-to-back, we have $\vec{P}=-\vec{K}$. Consequently,
we can write $K\cdot q_{\perp}=0,$ and the orthogonality condition
demands $K\cdot\epsilon=0$. The on-shell condition of photon $K^{2}=0$
is applied, and it can be verified that $P\cdot\epsilon=0$. The full
expressions of $b_{1,2}$ are given as 
\begin{equation}
b_{1,2}=\frac{-M_{Z}^{2}\pm\sqrt{(M_{Z}^{2}-M^{2})^{2}+4M^{2}\omega^{2}}}{2M}
\end{equation}
One readily finds that $b_{1}$ and $b_{2}$ correspond to a small
value and a large value, which are approximately $-M/2$ and $-M_{Z}^{2}/M$,
respectively. With the help of Eq.~\ref{eq:d} and Eq.~\ref{popd},
we can simplify Eq.~\ref{M14d} into: 
\begin{align}
\mathcal{M}_{1} & =i\frac{\sqrt{3}ee_{Q}gg_{v}^{Q}}{2\cos\theta_{W}}\int\frac{\mathrm{d}^{3}\vec{q}}{(2\pi)^{3}}\int\frac{\mathrm{d}q_{P}}{2\pi}\operatorname{Tr}\Biggl[\biggl(\frac{\Lambda_{1}^{+}(q_{\perp})}{q_{P}+\tfrac{1}{2}M-\omega+i\epsilon}+\frac{\Lambda_{1}^{-}(q_{\perp})}{q_{P}+\tfrac{1}{2}M+\omega-i\epsilon}\biggr)\bar{\Gamma}_{S}(q_{\perp})\nonumber \\
 & \quad\biggl(\frac{\Lambda_{2}^{+}(q_{\perp})}{-q_{P}+\tfrac{1}{2}M-\omega+i\epsilon}+\frac{\Lambda_{2}^{-}(q_{\perp})}{-q_{P}+\tfrac{1}{2}M+\omega-i\epsilon}\biggr)\slashed{\epsilon}_{\gamma}\frac{\bigl(\tfrac{1}{2}+q_{P}/M\bigr)\slashed{P}+\slashed{q}_{\perp}+\slashed{K}+m}{(q_{P}-b_{1}-i\epsilon)(q_{P}-b_{2}+i\epsilon)}\slashed{\varepsilon}_{Z}\Biggr]
\end{align}
With the Salpeter equations in Eq.~\ref{eq:4inde}, we can express
the quantity in trace bracket as 
\begin{align}
\mathcal{M}_{1} & =i\frac{\sqrt{3}ee_{Q}gg_{v}^{Q}}{2\cos\theta_{W}}\text{\ensuremath{\int\frac{\mathrm{d}^{3}\vec{q}}{(2\pi)^{3}}\int\frac{\mathrm{d}q_{P}}{2\pi}}Tr}\Biggl[\frac{(M-2\omega)\bar{\psi}_{\eta}^{++}(q_{\perp})\slashed{\epsilon}_{\gamma}\big((1/2+q_{P}/M)\slashed{P}+\slashed{q}_{\perp}+\slashed{K}+m\big)\slashed{\varepsilon}_{Z}}{(q_{P}+\frac{1}{2}M-\omega+i\epsilon)(-q_{P}+\frac{1}{2}M-\omega+i\epsilon)(q_{P}-b_{1}-i\epsilon)(q_{P}-b_{2}+i\epsilon)}\nonumber \\
 & -\frac{(M+2\omega)\bar{\psi}_{\eta}^{--}(q_{\perp})\slashed{\epsilon}_{\gamma}\big((1/2+q_{P}/M)\slashed{P}+\slashed{q}_{\perp}+\slashed{K}+m\big)\slashed{\varepsilon}_{Z}}{(q_{P}+\frac{1}{2}M+\omega-i\epsilon)(-q_{P}+\frac{1}{2}M+\omega-i\epsilon)(q_{P}-b_{1}-i\epsilon)(q_{P}-b_{2}+i\epsilon)}\Biggr]
\end{align}
Because of the last two Salpeter equations in Eq.~\ref{eq:4inde},
the $+-$ and $-+$ terms vanish \cite{Bhatnagar:2022wgg}. The quantities
$\psi_{\eta}^{++}(q_{\perp})$ and $\psi_{\eta}^{--}(q_{\perp})$
can then be obtained using Eq.~\ref{projwave}. 
Carrying out contour integration over the poles originating from each term's denominator in the complex $q_{P}$ plane by means of the residue theorem, we arrive at
\begin{align}
\mathcal{M}_{1} & =-\frac{\sqrt{3}ee_{Q}gg_{v}^{Q}}{2\cos\theta_{W}}\text{\ensuremath{\int\frac{\mathrm{d}^{3}\vec{q}}{(2\pi)^{3}}}Tr}\Biggl[C_{1}\bar{\psi}_{\eta}^{++}(q_{\perp})\slashed{\varepsilon}_{\gamma}\big((1-\frac{\omega}{M})\slashed{P}+\slashed{q}_{\perp}+\slashed{K}+m\big)\slashed{\epsilon}_{Z}\nonumber \\
 & +C_{2}\bar{\psi}_{\eta}^{--}(q_{\perp})\slashed{\varepsilon}_{\gamma}\big((-\frac{\omega}{M})\slashed{P}+\slashed{q}_{\perp}+\slashed{K}+m\big)\slashed{\epsilon}_{Z}\nonumber \\
 & +C_{3}\bar{\psi}_{\eta}^{--}(q_{\perp})\slashed{\varepsilon}_{\gamma}\big((1/2+b_{1}/M)\slashed{P}+\slashed{q}_{\perp}+\slashed{K}+m\big)\slashed{\epsilon}_{Z}\nonumber \\
 & +C_{4}\bar{\psi_{\eta}}^{++}(q_{\perp})\slashed{\varepsilon}_{\gamma}\big((1/2+b_{1}/M)\slashed{P}+\slashed{q}_{\perp}+\slashed{K}+m\big)\slashed{\epsilon}_{Z}\Biggr]\label{m1f}
\end{align}
One can verify that the results of the integration are the same regardless
of whether the contour is closed below or above the $q_{P}$-plane.
The full expressions for $C_{1},\dots,C_{4}$ are as follows: 
\begin{align}
C_{1}=-\frac{1}{(\frac{1}{2}M-\omega-b_{1})(\frac{1}{2}M-\omega-b_{2})},\, & C_{2}=-\frac{1}{(\frac{1}{2}M+\omega+b_{1})(\frac{1}{2}M+\omega+b_{2})}\label{C1}\nonumber \\
C_{3}=-\frac{(M+2\omega)}{((\frac{1}{2}M+\omega)^{2}-b_{1}^{2})(b_{1}-b_{2})},\, & C_{4}=-\frac{(M-2\omega)}{((\frac{1}{2}M-\omega)^{2}-b_{1}^{2})(b_{1}-b_{2})}
\end{align}
Here we can consider $C_{4}=0$ as $M\sim2\omega$, which leads to
\begin{align}
\mathcal{M}_{1} & =-\frac{\sqrt{3}ee_{Q}gg_{v}^{Q}}{2\cos\theta_{W}}\text{\ensuremath{\int\frac{\mathrm{d}^{3}\vec{q}}{(2\pi)^{3}}}Tr}\Biggl[C_{1}\bar{\psi}_{\eta}^{++}(q_{\perp})\slashed{\varepsilon}_{\gamma}\big((1-\frac{\omega}{M})\slashed{P}+\slashed{q}_{\perp}+\slashed{K}+m\big)\slashed{\epsilon}_{Z}\nonumber \\
 & +C_{2}\bar{\psi}_{\eta}^{--}(q_{\perp})\slashed{\varepsilon}_{\gamma}\big((-\frac{\omega}{M})\slashed{P}+\slashed{q}_{\perp}+\slashed{K}+m\big)\slashed{\epsilon}_{Z}\nonumber \\
 & +C_{3}\bar{\psi}_{\eta}^{--}(q_{\perp})\slashed{\varepsilon}_{\gamma}\big((1/2+b_{1}/M)\slashed{P}+\slashed{q}_{\perp}+\slashed{K}+m\big)\slashed{\epsilon}_{Z}\Biggr]\label{m1com}
\end{align}
It can be noticed that the expressions of $C_{i}$ are quite general,
with the input parameter being the quark mass $m$, which enters through
the $\omega=\sqrt{m_{i}^{2}+\vec{q}^{2}}$. The other quantities present
are the mass $M$ of the produced meson and the mass $M_{Z}$ of Z
boson. Thus, the $C_{i}$ are the functions of $|\vec{q}|$. To investigate
the continuity of the coefficients, we take the $\eta_{c}(1S)$ state
as example and plot their variations versus $|\vec{q}|$ in Fig.~\ref{fig:ceta}.
Discontinuities are observed for $C_{1}$ in the range $0<|\vec{q}|<4\,\text{GeV}$,
which exhibits a sharp, pole-like behavior. 
This kind of discontinuity arises from the coincidence of the fermion
propagator pole and the contour integration pole. Whereas $C_{2,3}$
are continuous over the entire range of $|\vec{q}|$.

After performing Dirac trace in Eq.~\ref{m1com}, we obtain 
\begin{equation}
\mathcal{M}_{1}=i\frac{\sqrt{3}ee_{Q}gg_{v}^{Q}}{2\cos\theta_{W}}\biggl[\xi_{1}\epsilon_{\mu\nu\sigma\rho}K^{\mu}P^{\nu}\varepsilon_{z}^{\sigma}\varepsilon_{\gamma}^{\rho}\biggr]\label{M1eta}
\end{equation}
In the above expression $\xi_{1}$ is the form factor, and its full
expression is given by 
\begin{align}
\xi_{1} & =\int\frac{\mathrm{d}^{3}\vec{q}}{(2\pi)^{3}}\frac{2\omega(C_{1}+C_{2}+C_{3})\phi_{1}+M(C_{1}-C_{2}-C_{3})\phi_{2}}{\omega}
\end{align}
Similarly, by employing the same steps we can simplify $\mathcal{M}_{2}$
as: 
\begin{align}
\mathcal{M}_{2} & =-\frac{\sqrt{3}ee_{Q}gg_{v}^{Q}}{2\cos\theta_{W}}\text{\ensuremath{\int\frac{\mathrm{d}^{3}\vec{q}}{(2\pi)^{3}}}Tr}\Biggl[C_{1}^{\prime}\bar{\psi}_{\eta}^{++}(q_{\perp})\slashed{\varepsilon}_{Z}\big((-\frac{\omega}{M})\slashed{P}+\slashed{q}_{\perp}+\slashed{K}+m\big)\slashed{\epsilon}_{\gamma}\nonumber \\
 & +C_{2}^{\prime}\bar{\psi}_{\eta}^{--}(q_{\perp})\slashed{\varepsilon}_{Z}\big((-1-\frac{\omega}{M})\slashed{P}+\slashed{q}_{\perp}+\slashed{K}+m\big)\slashed{\epsilon}_{\gamma}\label{m2com}\\
 & +C_{3}^{\prime}\bar{\psi}_{\eta}^{--}(q_{\perp})\slashed{\varepsilon}_{Z}\big((-1/2+d_{1}/M)\slashed{P}+\slashed{q}_{\perp}+\slashed{K}+m\big)\slashed{\epsilon}_{\gamma}\Biggr]\nonumber 
\end{align}
One can find that $d_{1}=-b_{2}$, $d_{2}=-b_{1}$ and the expressions
for $C_{1}^{\prime},C_{2}^{\prime}$, and $C_{3}^{\prime}$ are as
follows: 
\begin{align}
C_{1}^{\prime}=-\frac{1}{(\frac{1}{2}M-\omega-d_{1})(\frac{1}{2}M-\omega-d_{2})}, & C_{2}^{\prime}=-\frac{1}{(\frac{1}{2}M+\omega+d_{1})(\frac{1}{2}M+\omega+d_{2})}\label{Cc}\nonumber \\
C_{3}^{\prime}=-\frac{(M+2\omega)}{((\frac{1}{2}M+\omega)^{2}-d_{1}^{2})(d_{1}-d_{2})}
\end{align}

The plots for $C_{i}^{\prime}$ of $\eta_{c}$ are shown in Fig.~\ref{fig:ceta},
which are continuous over the range $0<|\vec{q}|<4\,\text{GeV}$.

It follows from the expressions of $C_{i}$ given in Eqs.~\ref{C1}
and ~\ref{Cc} that all $C_{i},\,C_{i}'$ carry dimensions of inverse
energy squared. Under the condition $M\ll M_{Z}$, each $C_{i},\,C_{i}'$
takes the asymptotic form 
\[
\frac{1}{M_{Z}^{2}}\left(\tilde{C}_{i}(\vec{q})+\mathcal{O}\!\left(\frac{M^{2}}{M_{Z}^{2}}\right)\right),
\]
while $C_{3}'$ is of order $\mathcal{O}\!\left(\dfrac{M^{2}}{M_{Z}^{4}}\right)$.
Here $\tilde{C}_{i}$ denote dimensionless parameters.

After taking Dirac trace of Eq.~\ref{m2com}, we obtain 
\begin{equation}
\mathcal{M}_{2}=i\frac{\sqrt{3}ee_{Q}gg_{v}^{Q}}{2\cos\theta_{W}}\biggl[\xi_{2}\epsilon_{\mu\nu\sigma\rho}K^{\mu}P^{\nu}\varepsilon_{z}^{\sigma}\varepsilon_{\gamma}^{\rho}\biggr]\label{M2 eta}
\end{equation}
Here, the form factor is 
\begin{align}
\xi_{2} & =\int\frac{\mathrm{d}^{3}\vec{q}}{(2\pi)^{3}}\frac{2\omega(C_{1}^{'}+C_{2}^{'}+C_{3}^{'})\phi_{1}+M(C_{1}^{'}-C_{2}^{'}-C_{3}^{'})\phi_{2}}{\omega}.
\end{align}
Finally, adding Eq.~\ref{M1eta} and Eq.~\ref{M2 eta}, we obtain
the total 3D-BS amplitude. 
\begin{equation}
\mathcal{M}_{Z\rightarrow\eta_{c}\gamma}=i\frac{\sqrt{3}ee_{Q}gg_{v}^{Q}}{2\cos\theta_{W}}\biggl[\xi\epsilon_{\mu\nu\sigma\rho}K^{\mu}P^{\nu}\varepsilon_{z}^{\sigma}\varepsilon_{\gamma}^{\rho}\biggr]\label{eta am}
\end{equation}
In the above expression, $\xi=\xi_{1}+\xi_{2}$. 
We observe that the relativistic correction terms cancel out, leaving
only the LO term remain in the amplitude, which shows LO effects dominate
in this process.


The two-body radiative decay width for $Z\rightarrow H_{Q\bar{Q}}+\gamma$
can be defined as follows 
\begin{equation}
\Gamma_{Z\rightarrow H_{Q\bar{Q}}\gamma}=\frac{1}{3}\cdot\frac{1}{2M_{Z}}\cdot\frac{M_{Z}^{2}-M^{2}}{8\pi M_{Z}^{2}}\cdot|\mathcal{M}|^{2}\label{def}
\end{equation}
The factor 1/3 comes from the Z-boson polarization averaging. Substitute
Eq.~\ref{eta am} into Eq.~\ref{def} and carry out the straightforward
mathematical steps. 
\begin{align}
\Gamma_{Z\rightarrow\eta_{c}\gamma} & =\frac{1}{2^{7}\pi}\biggl(\frac{ee_{Q}gg_{v}^{Q}\xi}{\cos\theta_{W}}\biggr)^{2}\frac{(M_{Z}^{2}-M^{2})^{3}}{M_{Z}^{3}}\label{eta dw}
\end{align}

With $M\ll M_{Z}$, we have 
\begin{align}
\Gamma_{Z\to\eta_{c}\gamma}\approx\frac{1}{2^{7}\,\pi}\biggl(\frac{e\,e_{Q}\,g\,g_{v}^{Q}}{\cos\theta_{W}}\biggr)^{\!2}\frac{M^{2}}{M_{Z}}\,\tilde{A}_{\eta_{c}}\label{eq:etadwapprox}
\end{align}
where $\tilde{A}_{\eta_{c}}=\frac{M_{Z}^{4}\xi^{2}}{M^{2}}$ is a
dimensionless parameter that is free of suppression or enhancement arising from powers of $M_{Z}$. The
decay width exhibits a leading $1/M_{Z}$ suppression due to the large
$Z$ boson mass, as seen in the prefactor of Eq.~\ref{eq:etadwapprox}.

\subsection{$1^{--}$ Charmonium}

Now, we study the vector S-wave meson process, $Z\rightarrow\psi(nS)+\gamma$
with $n=1,2,3$, and calculate their decay widths. To evaluate the
decay width we need to introduce BS wavefunction; therefore, we begin
with that. For the vector meson with $J^{PC}=1^{--}$, the general
decomposition of the BS wavefunction in terms of various Dirac structures
is given by \cite{Negash2019,Chang2010,Bhatnagar:2024ykb}: 
\begin{align}
\psi_{v}(q_{\perp}) & =q_{\perp}\cdot\varepsilon_{\perp}^{\lambda}\Bigl[\phi_{1}(q_{\perp})+\frac{\slashed{q}}{M}\phi_{3}(q_{\perp})+\frac{\slashed{P}\slashed{q}}{M}\phi_{4}(q_{\perp})\Bigr]+M\slashed{\varepsilon}_{\perp}^{\lambda}\phi_{5}(q_{\perp})\nonumber \\
 & +\slashed{\varepsilon}_{\perp}^{\lambda}\slashed{P}\phi_{6}(q_{\perp})+\Bigl[\frac{\slashed{P}\slashed{\varepsilon}_{\perp}^{\lambda}\slashed{q}_{\perp}}{M}-\frac{q_{\perp}\cdot\varepsilon_{\perp}^{\lambda}\slashed{P}}{M}\Bigr]\phi_{2}(q_{\perp})\label{eq:vecttt-1}
\end{align}

In the above expression, $M$ and $\varepsilon_{\perp}^{\lambda}$
represent the mass and polarization vector of vector quarkonium, respectively.
It is straightforward to see that it contains LO terms and relativistic
correction terms. Substituting Eq.~\ref{eq:vecttt-1} into the last
two equations of Eq.~\ref{eq:4inde} and evaluating the Dirac traces
on both sides yields the constraint conditions on the components of
the wave function: 
\begin{align}
\phi_{1}(q_{\perp})=\frac{-\phi_{3}(q_{\perp})\vec{q}^{2}+M^{2}\phi_{5}(q_{\perp})}{Mm}, & \phi_{2}(q_{\perp})=-\frac{M}{m}\phi_{6}(q_{\perp})
\end{align}

As a result, there are only four independent components left in the
wavefunction. 
\begin{align}
\psi_{v}(q_{\perp}) & =q_{\perp}\cdot\varepsilon_{\perp}^{\lambda}\Bigl(\frac{-\vec{q}^{2}}{Mm}+\frac{\slashed{q}}{M}\Bigr)\phi_{3}(q_{\perp})+q_{\perp}\cdot\varepsilon_{\perp}^{\lambda}\frac{\slashed{P}\slashed{q}}{M}\phi_{4}(q_{\perp})\nonumber \\
 & +\Bigl(q_{\perp}\cdot\varepsilon_{\perp}^{\lambda}\frac{M}{m}+M\slashed{\varepsilon}_{\perp}^{\lambda}\Bigr)\phi_{5}(q_{\perp})+\Bigl(\slashed{\varepsilon}_{\perp}^{\lambda}\slashed{P}-\frac{\slashed{P}\slashed{\varepsilon}_{\perp}^{\lambda}\slashed{q}_{\perp}}{m}+\frac{q_{\perp}\cdot\varepsilon_{\perp}^{\lambda}\slashed{P}}{m}\Bigr)\phi_{6}(q_{\perp})\label{eq:vettf-1}
\end{align}

It is important to notice that the wave function Eq.~\ref{eq:vettf-1}
involve S\textminus D wave mixing property. The components $\phi_{5}$
or $\phi_{6}$ are of S-wave nature, and represent the larger contributions,
while, the components proportional to $\phi_{3}$ or $\phi_{4}$ are
of D-wave nature suppressed by $v^2$ and are comparatively much smaller.
Therefore, we safely neglect them, following the treatments in previous BS studies \cite{Bhatnagar:2024ykb,Negash2019}.
Substituting the wave function into first two equations of Eq.~\ref{eq:4inde}
yields the coupled integral equations. 
\begin{align}
 & (M-2\omega)\biggl[\phi_{5}(q_{\perp})-\phi_{6}(q_{\perp})\frac{m}{\omega}\biggr]\nonumber \\
 & \quad=\int\frac{\mathrm{d}^{3}\vec{k}}{(2\pi)^{3}\omega^{2}}\biggl[\big(V_{s}+V_{v}\big)\phi_{5}(k_{\perp})\vec{k}\cdot\vec{q}-\big(V_{s}-V_{v}\big)\bigl(m^{2}\phi_{5}(k_{\perp})-\omega m\phi_{6}(k_{\perp})\bigr)\biggr],\\[6pt]
 & (M+2\omega)\biggl[\phi_{5}(q_{\perp})+\phi_{6}(q_{\perp})\frac{m}{\omega}\biggr]\nonumber \\
 & \quad=-\int\frac{\mathrm{d}^{3}\vec{k}}{(2\pi)^{3}\omega^{2}}\biggl[\big(V_{s}+V_{v}\big)\phi_{5}(k_{\perp})\vec{k}\cdot\vec{q}-\big(V_{s}-V_{v}\big)\bigl(m^{2}\phi_{5}(k_{\perp})+\omega m\phi_{6}(k_{\perp})\bigr)\biggr].
\end{align}
\label{eq:vect-coupled} We can solve the above equations numerically
for a given kernel to obtain the $\phi_{5}$ and $\phi_{6}$, as well
as mass spectrum, by treating it as an eigenvalue problem. The corresponding
normalization condition reads 
\[
4M^{2}\int\frac{\mathrm{d}^{3}\vec{q}}{(2\pi)^{3}}\frac{\phi_{5}\phi_{6}\omega}{m}=P^{0}.
\]

\begin{figure}[h]
\centering \includegraphics[scale=0.4]{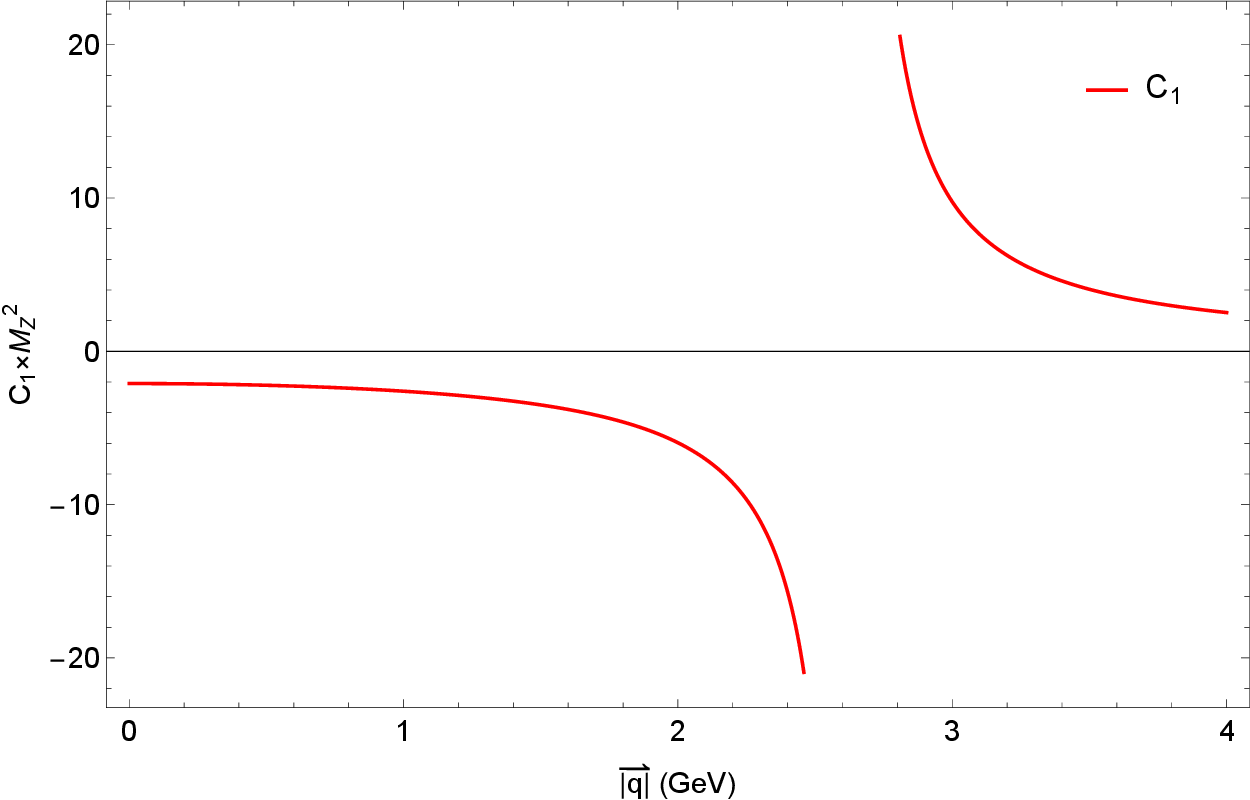}\includegraphics[scale=0.4]{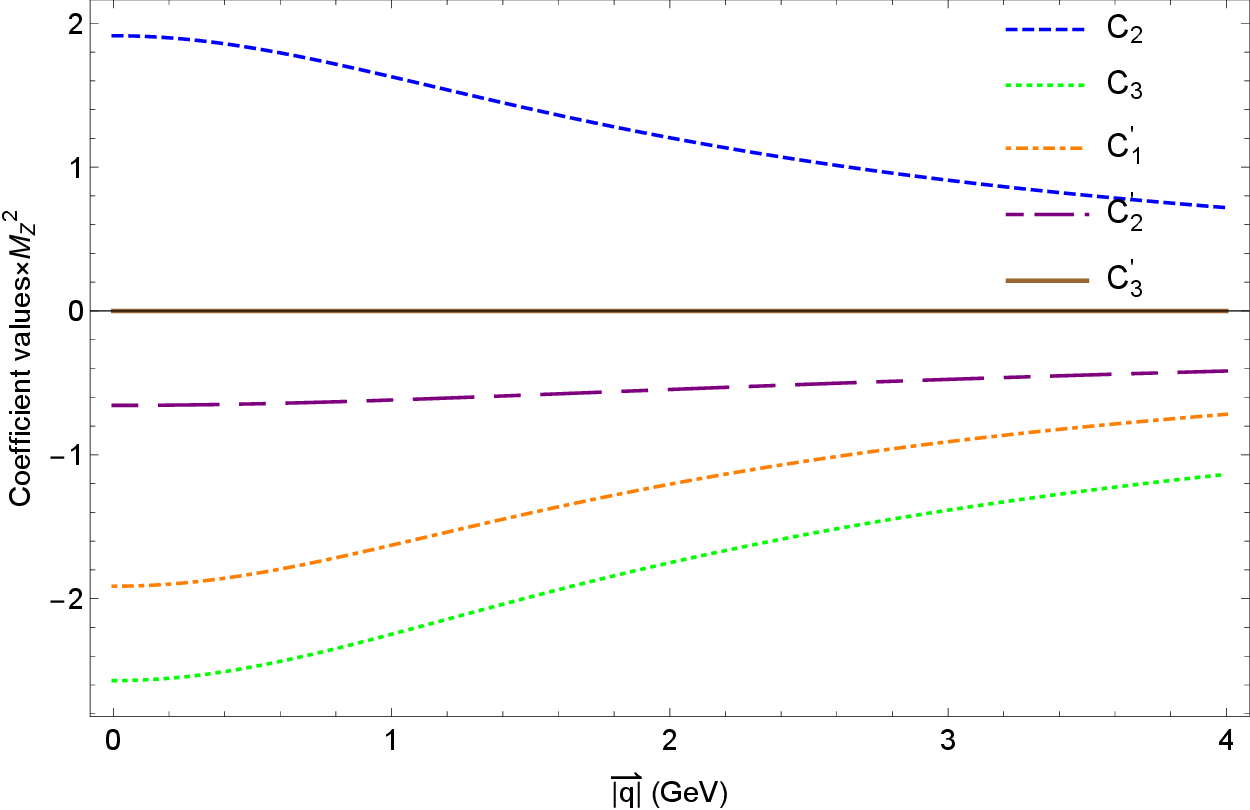}
\caption{ Coefficients $C_{1}$ (left) and $C_{i},\,C_{i}'$ (right) multiplied
by $M_{Z}^{2}$ versus $|\vec{q}|$ in the range $0<|\vec{q}|<4\,\text{GeV}$
for the $J/\psi$ state. }
\label{fig:cphi} 
\end{figure}

Now we are ready to evaluate amplitude for $Z\rightarrow\psi+\gamma$.
Similar to the case of $Z\rightarrow\eta_{c}+\gamma$, the corresponding
4D-BS amplitude can be reduced into 3D-BS amplitudes for both diagrams
as 
\begin{align}
\mathcal{M}_{1} & =\frac{\sqrt{3}ee_{Q}gg_{a}^{Q}}{2\cos\theta_{W}}\text{\ensuremath{\int\frac{\mathrm{d}^{3}\vec{q}}{(2\pi)^{3}}}Tr}\Biggl[C_{1}\bar{\psi}_{v}^{++}(q_{\perp})\slashed{\varepsilon}_{\gamma}\big((1-\frac{\omega}{M})\slashed{P}+\slashed{q}_{\perp}+\slashed{K}+m\big)\slashed{\epsilon}_{Z}\nonumber \\
 & +C_{2}\bar{\psi}_{v}^{--}(q_{\perp})\slashed{\varepsilon}_{\gamma}\big((-\frac{\omega}{M})\slashed{P}+\slashed{q}_{\perp}+\slashed{K}+m\big)\slashed{\epsilon}_{Z}+C_{3}\bar{\psi}_{v}^{--}(q_{\perp})\slashed{\varepsilon}_{\gamma}\big((1/2+b_{1}/M)\slashed{P}+\slashed{q}_{\perp}+\slashed{K}+m\big)\slashed{\epsilon}_{Z}\Biggr]\label{M1V}
\end{align}
and 
\begin{align}
\mathcal{M}_{2} & =\frac{\sqrt{3}ee_{Q}gg_{a}^{Q}}{2\cos\theta_{W}}\text{\ensuremath{\int\frac{\mathrm{d}^{3}\vec{q}}{(2\pi)^{3}}}Tr}\Biggl[C_{1}^{\prime}\bar{\psi}_{v}^{++}(q_{\perp})\slashed{\varepsilon}_{Z}\big((-\frac{\omega}{M})\slashed{P}+\slashed{q}_{\perp}+\slashed{K}+m\big)\slashed{\epsilon}_{\gamma}\nonumber \\
 & +C_{2}^{\prime}\bar{\psi}_{v}^{--}(q_{\perp})\slashed{\varepsilon}_{Z}\big((-1-\frac{\omega}{M})\slashed{P}+\slashed{q}_{\perp}+\slashed{K}+m\big)\slashed{\epsilon}_{\gamma}+C_{3}^{\prime}\bar{\psi}_{v}^{--}(q_{\perp})\slashed{\varepsilon}_{Z}\big((-1/2+d_{1}/M)\slashed{P}+\slashed{q}_{\perp}+\slashed{K}+m\big)\slashed{\epsilon}_{\gamma}\Biggr]\label{M2v}
\end{align}
The coefficients $C_{i}$ and $C_{i}^{'}$ are defined in Eq.~\ref{C1}
and Eq.~\ref{Cc}, respectively. 
Similar to the $0^{-+}$ states case, the coefficient $C_{1}$ exhibits a discontinuity, while 
the remaining coefficients are smooth over the integration range $0<|\vec{q}|<4\,\text{GeV}$.
This behavior is illustrated in Fig.~\ref{fig:cphi}, with $J/\psi$
taken as an example.

We can obtain $\psi_{v}^{++}(q_{\perp})$ and $\psi_{v}^{--}(q_{\perp})$
using Eq.~\ref{projwave}. Substituting them into Eq.~\ref{M1V}
and Eq.~\ref{M2v}, taking the Dirac trace, and summing the results
yields: 
\begin{align}
\mathcal{M}_{Z\rightarrow\psi\gamma} & =-i\frac{\sqrt{3}ee_{Q}gg_{a}^{Q}}{2\cos\theta_{W}}\biggl[J_{1}\epsilon^{K\epsilon_{z}\epsilon_{1}\epsilon_{\gamma}}+J_{2}\epsilon^{P\epsilon_{z}\epsilon_{1}\epsilon_{\gamma}}+J_{3}\epsilon^{I\epsilon_{z}\epsilon_{1}\epsilon_{\gamma}}\nonumber \\
 & \quad+J_{4}\bigl(I\cdot\epsilon_{z}\epsilon^{KP\epsilon_{1}\epsilon_{\gamma}}+I\cdot\epsilon_{\gamma}\epsilon^{KP\epsilon_{z}\epsilon_{1}}-P\cdot\epsilon_{z}\epsilon^{KI\epsilon_{1}\epsilon_{\gamma}}-\epsilon_{z}\cdot\epsilon_{1}\epsilon^{KPI\epsilon_{\gamma}}+\epsilon_{1}\cdot\epsilon_{\gamma}\epsilon^{KPI\epsilon_{z}}\bigr)\nonumber \\
 & \quad-J_{5}I\cdot\epsilon_{\gamma}\epsilon^{PI\epsilon_{z}\epsilon_{1}}-J_{6}I\cdot\epsilon_{1}\epsilon^{PI\epsilon_{z}\epsilon_{\gamma}}+\epsilon^{PI\epsilon_{1}\epsilon_{\gamma}}(J_{7}K\cdot\epsilon_{z}+J_{8}P\cdot\epsilon_{z}-J_{9}I\cdot\epsilon_{z})\nonumber \\
 & \quad+J_{10}\epsilon^{KPI\epsilon_{1}}\epsilon_{z}\cdot\epsilon_{\gamma}+J_{11}I\cdot\epsilon_{1}\epsilon^{KI\epsilon_{z}\epsilon_{\gamma}}\biggr]\label{eq:vector_amplitude}
\end{align}
Here, we have replaced effective 3D variable by $q_{\perp}^{\nu}=|\vec{q}|I^{\nu}$
where $I^{\nu}$ is a spatial unit vector along the direction of $q_{\perp}^{\nu}$,
and $\epsilon_{1}$ denotes polarization vector of $J/\psi$. The
full expression of the form factors $J_{i}$ are given by: 
\begin{align}
J_{1} & =2\frac{M}{m}\int\frac{\mathrm{d}^{3}\vec{q}}{(2\pi)^{3}}\big[(C_{1}+C_{1}'+C_{2}+C_{2}'+C_{3}+C_{3}')m\phi_{5}-(C_{1}+C_{1}'-C_{2}-C_{2}'-C_{3}-C_{3}')\omega\phi_{6}\big]\label{eq:ji}\nonumber \\
J_{2} & =\int\frac{\mathrm{d}^{3}\overrightarrow{q}}{(2\pi)^{3}}\frac{1}{m\omega}\biggl\{\Bigl[2\omega C_{1}(M-2\omega)+4C_{1}'\vec{q}^{2}+2C_{2}'(2m^{2}+\omega M)+\omega C_{3}(2b_{1}+M+2\omega)\nonumber \\
 & +C_{3}'(2m^{2}-2d_{1}\omega+M\omega-2\vec{q}^{2})\Bigr]m\phi_{5}+\Bigl[2\omega C_{1}(2\omega-M)-4C_{1}'\vec{q}^{2}+2C_{2}'(2m^{2}+\omega M)\nonumber \\
 & +\omega C_{3}(2b_{1}+M+2\omega)+C_{3}'(2m^{2}-2d_{1}\omega+M\omega-2\vec{q}^{2})\Bigr]\omega\phi_{6}\biggr\}\nonumber \\
J_{3} & =\int\frac{\mathrm{d}^{3}\vec{q}}{(2\pi)^{3}}\frac{|\vec{q}|}{m\omega}\biggl\{\Bigl[C_{1}(4M\omega-M^{2}-M_{Z}^{2})+C_{1}'(M^{2}-4M\omega-M_{Z}^{2})+C_{2}(M_{Z}^{2}-M^{2})\nonumber \\
 & +C_{2}'(M^{2}+M_{Z}^{2})+C_{3}(M_{Z}^{2}+2b_{1}M+2M\omega)+C_{3}'(M_{Z}^{2}-2d_{1}M-2M\omega)\Bigr]m\phi_{5}\nonumber \\
 & +\Bigl[C_{1}(M_{Z}^{2}+M^{2}-4M\omega)+C_{1}'(M_{Z}^{2}-M^{2}+4M\omega)+C_{2}(M_{Z}^{2}-M^{2})\nonumber \\
 & +C_{2}'(M^{2}+M_{Z}^{2})+C_{3}(2M\omega+M_{Z}^{2}+2b_{1}M)+C_{3}'(M_{Z}^{2}-2d_{1}M-2M\omega)\Bigr]\omega\phi_{6}\biggr\}\nonumber \\
J_{4} & =2\int\frac{\mathrm{d}^{3}\vec{q}}{(2\pi)^{3}}\frac{|\vec{q}|}{m\omega}\biggl[m(C_{1}+C_{1}'-C_{2}-C_{2}'-C_{3}-C_{3}')\phi_{5}-(C_{1}+C_{1}'+C_{2}+C_{2}'+C_{3}+C_{3}')\omega\phi_{6}\biggr]\nonumber 
\\
J_{5} & =4\int\frac{\mathrm{d}^{3}\vec{q}}{(2\pi)^{3}}\frac{\vec{q}^{2}}{m\omega}\biggl[(C_{1}-C_{2}-C_{3})m\phi_{5}-(C_{1}+C_{2}+C_{3})\omega\phi_{6}\biggr]\nonumber \\
J_{6} & =\int\frac{\mathrm{d}^{3}\vec{q}}{(2\pi)^{3}}\frac{\vec{q}^{2}}{m\omega}\biggl\{4\bigl(C_{1}'-C_{2}'-C_{3}'\bigr)m\phi_{5}+\Bigl[2\omega(2C_{1}-2C_{1}'+C_{3}-C_{3}')\nonumber \\
 & -M(2C_{1}-2C_{2}'-C_{3}-C_{3}')+2b_{1}C_{3}-2C_{3}'d_{1}\Bigr]\phi_{6}\biggr\}\nonumber \\
J_{7} & =J_{10}=2\int\frac{\mathrm{d}^{3}\vec{q}}{(2\pi)^{3}}\frac{|\vec{q}|}{m\omega}\biggl[(C_{1}-C_{1}'-C_{2}+C_{2}'-C_{3}+C_{3}')m\phi_{5}-(C_{1}-C_{1}'+C_{2}-C_{2}'+C_{3}-C_{3}')\omega\phi_{6}\biggr]\nonumber 
\\
J_{8} & =2\int\frac{\mathrm{d}^{3}\vec{q}}{(2\pi)^{3}}\frac{|\vec{q}|}{mM\omega}\biggl\{\Bigl[2C_{2}'(\omega+M)-2C_{1}'\omega+C_{3}'(M-2d_{1})\Bigr]m\phi_{5}\nonumber \\
 & +\Bigl[2C_{2}'(\omega+M)+2C_{1}'\omega+C_{3}'(M-2d_{1})\Bigr]\omega\phi_{6}\biggr\}\nonumber \\
J_{9} & =4\int\frac{\mathrm{d}^{3}\vec{q}}{(2\pi)^{3}}\frac{\vec{q}^{2}}{m\omega}\biggl[(C_{2}'-C_{1}'+C_{3}')m\phi_{5}+(C_{1}'+C_{2}'+C_{3}')\omega\phi_{6}\biggr]\nonumber \\
J_{11} & =2M\int\frac{\mathrm{d}^{3}\vec{q}}{(2\pi)^{3}}\frac{\vec{q}^{2}}{\omega m}\biggl[(C_{1}+C_{1}'-C_{2}-C_{2}'-C_{3}-C_{3}')\phi_{6}\biggr]
\end{align}

Relativistic correction to form factors can be incorporated order
by order in the expansion. One can observe that the amplitude (and
consequently the form factors) is expanded up to the second order
in $\vec{q}$. The above form factors can be calculated numerically,
and following traditional steps we can evaluate corresponding decay
width as 
\begin{align}
\Gamma_{Z\rightarrow\psi\gamma} & =\frac{1}{2^{8}\pi}\biggl(\frac{ee_{Q}gg_{a}^{Q}}{\cos\theta_{W}}\biggr)^{2}\frac{(M_{Z}^{2}-M^{2})}{M^{2}M_{Z}^{5}}A_{\psi}.\label{eq:psiwidth}
\end{align}
For brevity, we have introduced $A_{\psi}$ in the above expression,
and its full expression is given in the Appendix A.

Under the condition $M\ll M_{Z}$, we substitute Eq.~\ref{eq:Avapprox}
from Appendix A into the expression for the decay width and obtain
\begin{equation}
\Gamma_{Z\rightarrow\psi\gamma}\approx\frac{1}{2^{10}\pi}\biggl(\frac{e\,e_{Q}\,g\,g_{a}^{Q}}{\cos\theta_{\text{W}}}\biggr)^{\!2}M_{Z}\,\tilde{A}_{\psi}\label{eq:psiwithapprox}
\end{equation}
where $\tilde{A}_{\psi}=\dfrac{M_{Z}^{4}\,J_{4}^{2}}{M^{2}}$. Similar
to $\tilde{A}_{\eta_{c}}$, this factor is dimensionless and free
of $M_{Z}^{n}$ suppression or enhancement. The decay width is enhanced by an explicit
$M_{Z}$ factor.

\subsection{$0^{++}$ Charmonium}

\begin{figure}[h]
\centering \includegraphics[scale=0.4]{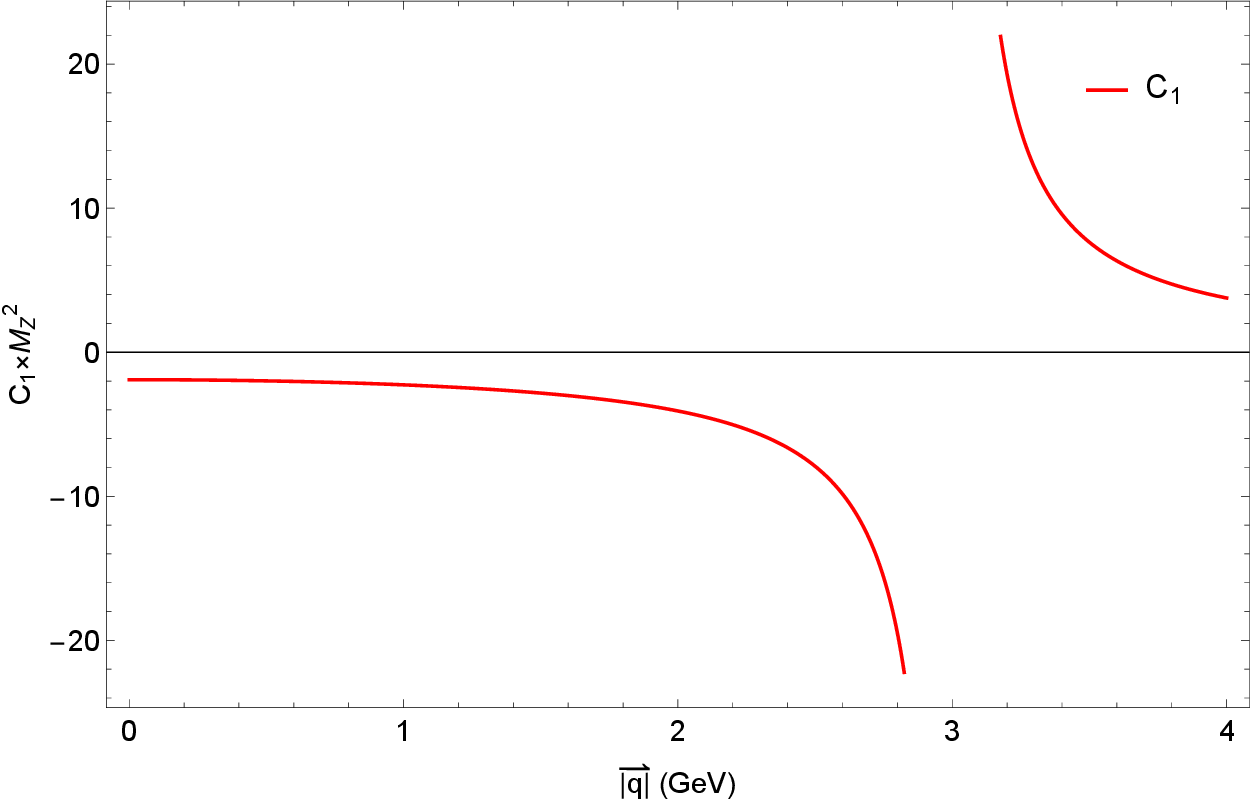}\includegraphics[scale=0.4]{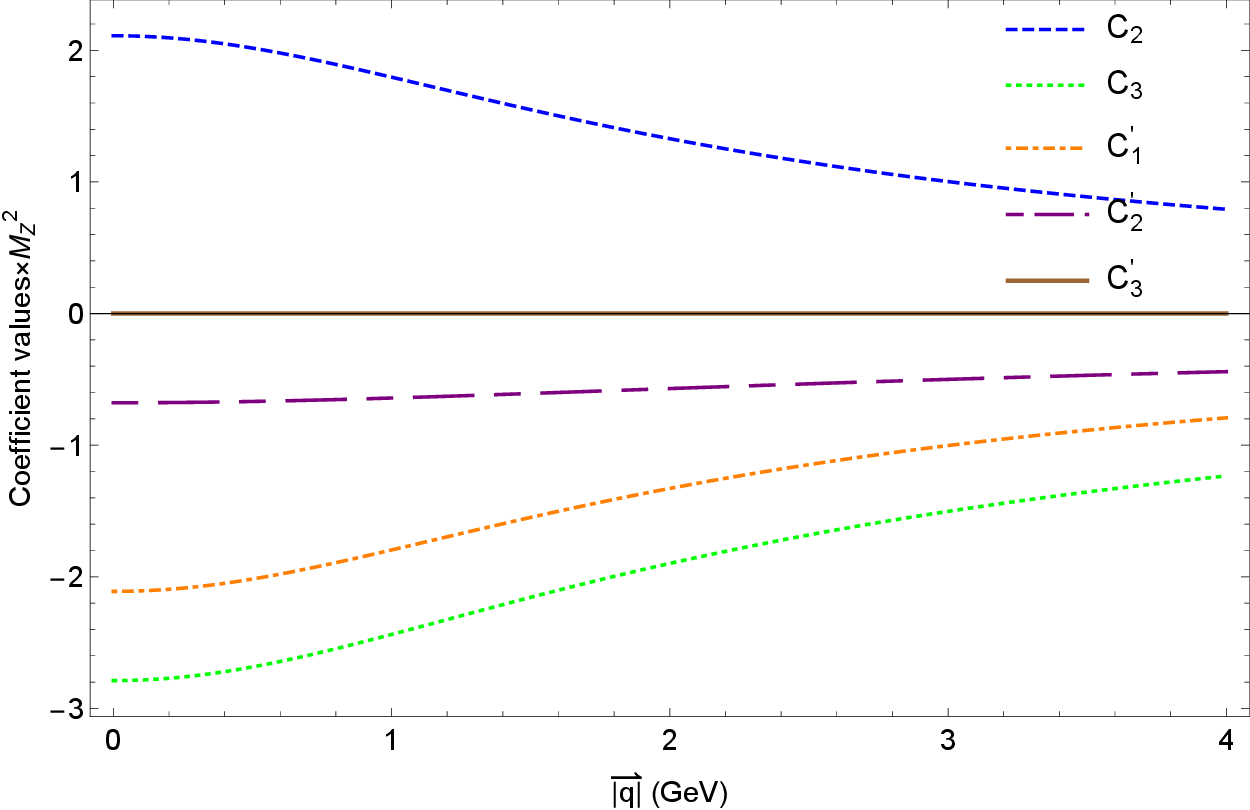}
\caption{ Coefficients $C_{1}$ (left) and $C_{i},\,C_{i}'$ (right) multiplied
by $M_{Z}^{2}$ versus $|\vec{q}|$ in the range $0<|\vec{q}|<4\,\text{GeV}$
for the $\chi_{c0}(1P)$ state. }
\label{fig:cchic0} 
\end{figure}

Now, we evaluate decay widths for $Z\rightarrow\chi_{c0}(mP)+\gamma$
with $m=1,2$. 
Before proceeding to the amplitude, we first introduce the corresponding
BS wavefunction. For the state with $J^{P}=0^{++}$, the general form
of the BS wave functions can be written as \cite{Chang2010,Wang:2007av}
\begin{equation}
\psi_{\chi_{c0}}(q_{\perp})=\phi_{1}(q_{\perp})\slashed{q}_{\perp}+\phi_{2}(q_{\perp})\frac{\slashed{P}\cdot\slashed{q}_{\perp}}{M}+\phi_{3}(q_{\perp})M.
\end{equation}
Here, $P$ and $M$ denote momentum and mass of the $\chi_{0}$, respectively.
The last two equations of Eq.~\ref{eq:4inde} impose constraints
on wavefunction as follows 
\begin{equation}
\phi_{3}(q_{\perp})=-\phi_{1}(q)\frac{\vec{q}^{2}}{Mm}.
\end{equation}
Therefore, the wavefunction takes the form 
\begin{equation}
\psi_{\chi_{c0}}(q_{\perp})=\Bigl(\slashed{q}_{\perp}-\frac{\vec{q}^{2}}{m}\Bigr)\phi_{1}(q_{\perp})+\frac{\slashed{P}\cdot\slashed{q}_{\perp}}{M}\phi_{2}(q_{\perp}).
\end{equation}
The corresponding normalization condition of wavefunction is given
by: 
\begin{equation}
4\int\frac{\mathrm{d}^{3}\vec{q}}{(2\pi)^{3}}\frac{\phi_{1}(q_{\perp})\phi_{2}(q_{\perp})\omega\vec{q}^{2}}{m}=P^{0}.
\end{equation}
Now, with the help of first two equations of Eq.~\ref{eq:4inde},
we obtain the coupled integral equations for the $\chi_{c0}$ states:
\begin{align}
(M-2\omega)\Bigl[\phi_{1}(q_{\perp})+\phi_{2}(q_{\perp})\frac{m}{\omega}\Bigr] & =\int\frac{\mathrm{d}^{3}\vec{k}}{(2\pi)^{3}\vec{q}^{2}}\frac{1}{\omega^{2}}\nonumber \\
 & \biggl\{(V_{s}+V_{v})\phi_{1}(k_{\perp})\vec{k}^{2}\vec{q}^{2}-(V_{s}-V_{v})\left[\phi_{1}(k_{\perp})m^{2}+\phi_{2}(k_{\perp})\omega m\right](\vec{k}\cdot\vec{q})\biggr\}\\
(M+2\omega)\Bigl[\phi_{1}(q_{\perp})-\phi_{2}(q_{\perp})\frac{m}{\omega}\Bigr] & =-\int\frac{\mathrm{d}^{3}\vec{k}}{(2\pi)^{3}\vec{q}^{2}}\frac{1}{\omega^{2}}\nonumber \\
 & \biggl\{(V_{s}+V_{v})\phi_{1}(k_{\perp})\vec{k}^{2}\vec{q}^{2}-(V_{s}-V_{v})\left[\phi_{1}(k_{\perp})m^{2}-\phi_{2}(k_{\perp})\omega m\right](\vec{k}\cdot\vec{q})\biggr\}\label{eq:0PP}
\end{align}
Now we are ready to proceed with the amplitude for the process. Replacing
$\psi_{\chi_{c0}}^{++}(q_{\perp})$ and $\psi_{\chi_{c0}}^{--}(q_{\perp})$
with $\psi_{\eta}^{++}(q_{\perp})$ and $\psi_{\eta}^{--}(q_{\perp})$
in Eq.~\ref{m1com} and Eq.~\ref{m2com}, respectively, carrying
out Dirac trace of each amplitude, summing them, and defining a unit
vector $I$ along the direction of $q_{\perp}$, we can obtain the
total amplitude via the vector coupling. 
\begin{align}
\mathcal{M}_{Z\rightarrow\chi_{c0}\gamma} & =-\frac{\sqrt{3}ee_{Q}gg_{v}^{Q}}{2\cos\theta_{W}}\biggl[I\cdot\epsilon_{\gamma}(X_{1}K\cdot\epsilon_{z}+X_{2}I\cdot\epsilon_{z}+X_{3}P\cdot\epsilon_{z})\biggr].
\end{align}
The full expression of the form factors $X_{i}$ are given below.
\begin{align}
X_{1}= & 2\int\frac{\mathrm{d}^{3}\vec{q}}{(2\pi)^{3}}\frac{|\vec{q}|}{\omega}\left[(C_{1}-C_{1}^{'}-C_{2}+C_{2}^{'}-C_{3}+C_{3}^{'})m\phi_{2}+(C_{1}-C_{1}^{'}+C_{2}-C_{2}^{'}+C_{3}-C_{3}^{'})\omega\phi_{1}\right]\nonumber 
\\
X_{2}= & 4\int\frac{\mathrm{d}^{3}\vec{q}}{(2\pi)^{3}}\frac{\vec{q}^{2}}{\omega}\biggl[(C_{1}^{'}+C_{2}+C_{2}^{'}+C_{3}+C_{3}^{'})\omega\phi_{1}+(C_{1}^{'}-C_{2}-C_{2}^{'}-C_{3}-C_{3}^{'})m\phi_{2}+C_{1}(m\phi_{2}+\omega\phi_{1})\biggr]\nonumber \\
X_{3}= & \int\frac{\mathrm{d}^{3}\vec{q}}{(2\pi)^{3}}\frac{|\vec{q}|}{\omega M}\biggl\{\Bigl[2C_{1}(M-2\omega)-2C_{2}'(M+2\omega)+C_{3}(M+2\omega+2b_{1})-C_{3}'(M+2\omega-2d_{1})\Bigr]\omega\phi_{1}\nonumber \\
 & +\Bigl[2C_{1}(M-2\omega)+2C_{2}'(M+2\omega)-C_{3}(M+2\omega+2b_{1})+C_{3}'(M+2\omega-2d_{1})\Bigr]m\phi_{2}\biggr\}
\end{align}
We can observe that the amplitude is expanded up to the second order
in $|\vec{q}|$. 
The behaviors of the coefficients $C_{i}$ and $C_{i}^{'}$ are shown in Fig.~\ref{fig:cchic0},
with $\chi_{c0}(1P)$ taken as an example.

The corresponding decay width reads as 
\begin{align}
\Gamma_{Z\rightarrow\chi_{c0}\gamma} & =\frac{1}{2^{8}\pi}\biggl(\frac{ee_{Q}gg_{v}^{Q}}{\cos\theta_{W}}\biggr)^{2}\frac{(M_{Z}^{2}-M^{2})}{M_{Z}^{5}}\Bigl[4M_{Z}^{2}X_{2}^{2}+(X_{1}-X_{3})^{2}(M_{Z}^{2}-M^{2})^{2}\Bigr]\label{eq:dwchic0}
\end{align}

From the order analysis of $C_{i}$, we find that dimensionless $X_{1,3}$
are of order $R_{1}/M_{Z}^{2}$. Whereas $X_{2}$ carries mass-energy
dimensions and is of order $R_{2}/M_{Z}^{2}$. Here the notation $R_{n}$
is defined as 
\begin{equation}
R_{n}\equiv\int\vec{q}^{n}\,\phi\,\tilde{C}_{i}(\vec{q})\,\mathrm{d}^{3}\vec{q}.\label{eq:Rn}
\end{equation}
Therefore, the two terms in the sum in the decay width expression
(Eq.~\ref{eq:dwchic0}) are of order $\dfrac{R_{2}^{2}}{M_{Z}^{5}}$
and $\dfrac{R_{1}^{2}}{M_{Z}^{3}}$, respectively. The large mass
of the Z boson strongly suppresses the decay width. In addition, the
negative sign between $X_{1,3}$ in the second term leads to partial
cancellation of their contributions, further lowering the overall
magnitude.

\subsection{$1^{++}$ Charmonium}

Similarly, we evaluate the decay widths for $Z\rightarrow\chi_{c1}(mP)+\gamma$
with $m=1,2$.
To evaluate the decay width, we need to introduce BS wavefunction;
therefore, we begin with that. For the $J^{PC}=1^{++}$ state, the
general decomposition of BS wave function in terms of various Dirac
structures is given by \cite{Geng2019,Chang2010} 
\begin{equation}
\psi_{\chi_{c1}}(q_{\perp})=i\frac{\varepsilon_{\mu\nu\alpha\beta}P^{\nu}q_{\perp}^{\alpha}\epsilon^{\beta}}{M^{2}}\gamma^{\mu}\Bigl[\phi_{1}(q_{\perp})M+\phi_{2}(q_{\perp})\slashed{P}+\phi_{3}(q_{\perp})\slashed{P}\slashed{q}/M\Bigr]
\end{equation}
Here, $P$ and $\text{\text{\ensuremath{\epsilon}}}$ denote the momentum
and polarization vector of the $\chi_{c1}$ with mass $M$, respectively.
The last two equations of Eq.~\ref{eq:4inde} impose constraints
on wavefunction. 
\begin{equation}
\phi_{3}(q)=\phi_{2}(q_{\perp})\frac{M}{m},
\end{equation}
that leads to 
\begin{equation}
\psi_{\chi_{c1}}(q_{\perp})=i\frac{\varepsilon_{\mu\nu\alpha\beta}P^{\nu}q_{\perp}^{\alpha}\epsilon^{\beta}}{M^{2}}\gamma^{\mu}\Bigl[\phi_{1}(q_{\perp})M+\phi_{2}(q_{\perp})(\slashed{P}+\slashed{P}\slashed{q}/m)\Bigr].
\end{equation}
The normalization reads as 
\begin{equation}
8\int\frac{\mathrm{d}^{3}\vec{q}}{(2\pi)^{3}}\frac{\phi_{1}(q_{\perp})\phi_{2}(q_{\perp})\omega\vec{q}^{2}}{3m}=P^{0}.
\end{equation}

\begin{figure}[h]
\centering \includegraphics[scale=0.4]{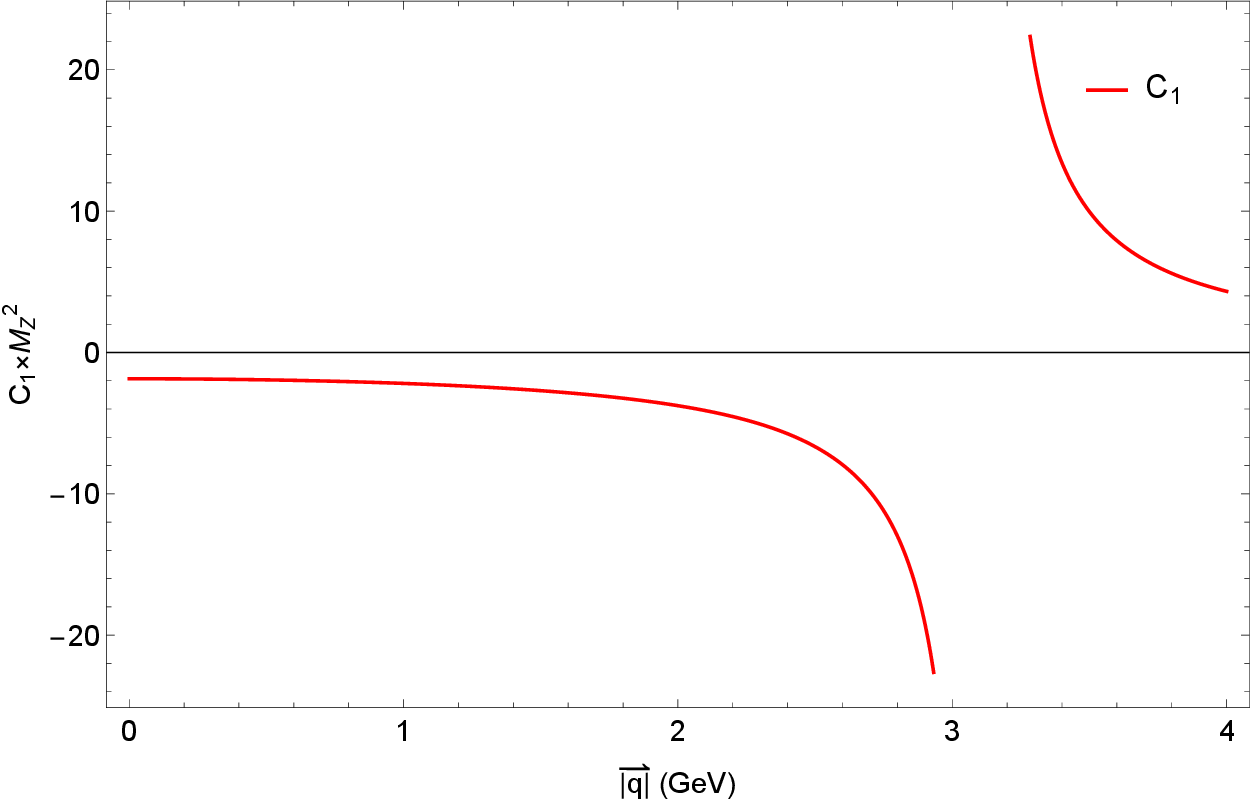}\includegraphics[scale=0.4]{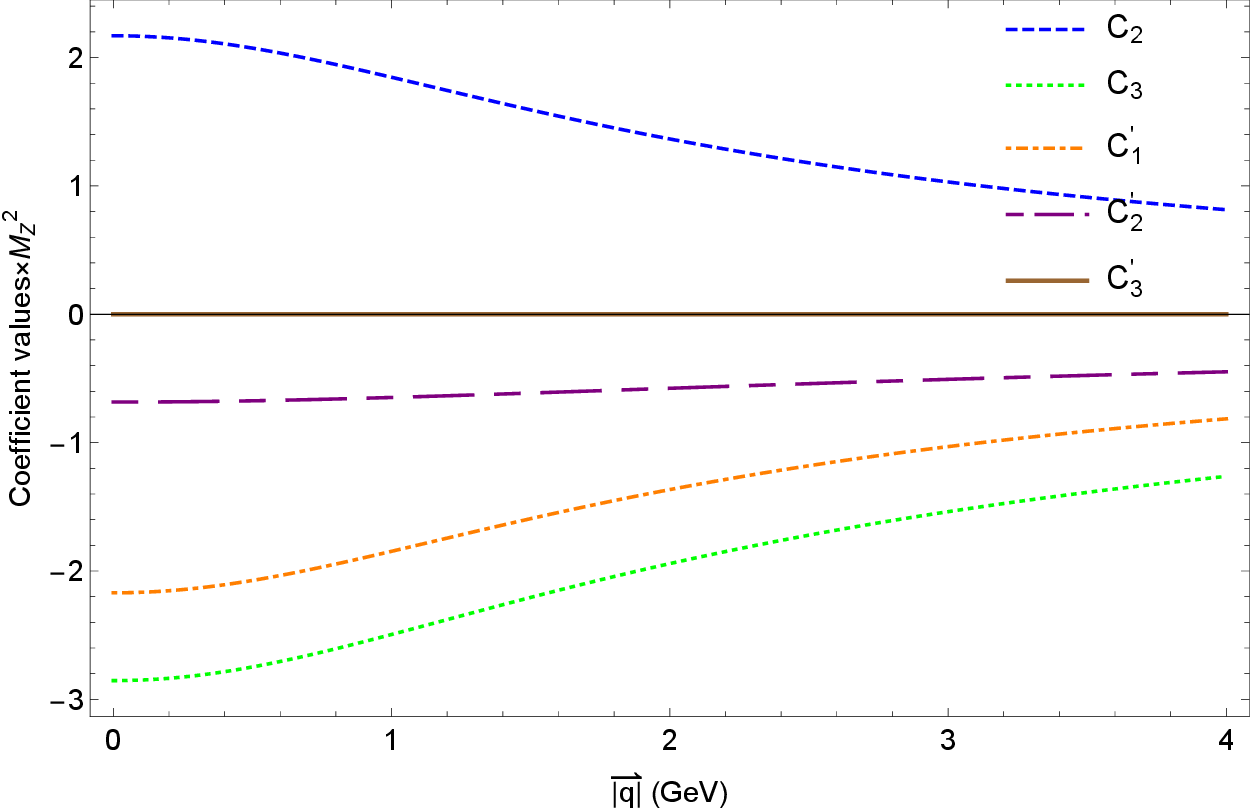}
\caption{ Coefficients $C_{1}$ (left) and $C_{i},\,C_{i}'$ (right) multiplied
by $M_{Z}^{2}$ versus $|\vec{q}|$ in the range $0<|\vec{q}|<4\,\text{GeV}$
for the $\chi_{c1}(1P)$ state. }
\label{fig:cchic1} 
\end{figure}

With the help of first two equations of Eq.~\ref{eq:4inde}, we obtain
the corresponding coupled integral equations as follows 
\begin{align}
(M-2\omega)\Bigl[\phi_{1}(q_{\perp})-\phi_{2}(q_{\perp})\frac{\omega}{m}\Bigr] & =-\int\frac{\mathrm{d}^{3}\vec{k}}{(2\pi)^{3}}\frac{1}{2\omega m\vec{q}^{2}}\nonumber \\
 & \biggl[(V_{s}+V_{v})\phi_{2}(k_{\perp})\Bigl[\vec{k}^{2}\vec{q}^{2}+(\vec{k}\cdot\vec{q})^{2}\Bigr]\nonumber \\
 & +2m(V_{s}-V_{v})\left[\phi_{1}(k_{\perp})\omega-\phi_{2}(k_{\perp})m\right](\vec{k}\cdot\vec{q})\biggr],
\end{align}
\begin{align}
(M+2\omega)\Bigl[\phi_{1}(q_{\perp})+\phi_{2}(q_{\perp})\frac{\omega}{m}\Bigr] & =-\int\frac{\mathrm{d}^{3}\vec{k}}{(2\pi)^{3}}\frac{1}{2\omega m\vec{q}^{2}}\nonumber \\
 & \biggl[(V_{s}+V_{v})\phi_{2}(k_{\perp})\Bigl[\vec{k}^{2}\vec{q}^{2}+(\vec{k}\cdot\vec{q})^{2}\Bigr]\nonumber \\
 & -2m(V_{s}-V_{v})\left[\phi_{1}(k_{\perp})\omega+\phi_{2}(k_{\perp})m\right](\vec{k}\cdot\vec{q})\biggr].\label{x1}
\end{align}
Now we proceed to the amplitude. Similar to the case of $Z\rightarrow\chi_{c0}\gamma$
, we substitute $\psi_{\chi_{c1}}^{++}(q_{\perp})$ and $\psi_{\chi_{c1}}^{--}(q_{\perp})$
into Eq.~\ref{m1com} and Eq.~\ref{m2com}. following the same calculation
procedures, we can obtain the total amplitude. 
\begin{align}
\mathcal{M}_{Z\rightarrow\chi_{c1}\gamma} & =-i\frac{\sqrt{3}ee_{Q}gg_{v}^{Q}}{2\cos\theta_{W}}\biggl[t_{1}I\cdot\epsilon_{\gamma}\epsilon^{IP\epsilon_{z}\epsilon_{1}}+\epsilon^{IP\epsilon_{1}\epsilon_{\gamma}}(t_{2}K\cdot\epsilon_{z}+t_{3}I\cdot\epsilon_{z}+t_{4}P\cdot\epsilon_{z})\nonumber \\
 & +2\epsilon^{KIP\epsilon_{1}}(t_{5}P\cdot\epsilon_{z}I\cdot\epsilon_{\gamma}+t_{6}\epsilon_{\gamma}\cdot\epsilon_{z})\biggr]
\end{align}
Here, $\epsilon_{1}$ is polarization vector of $\chi_{c1}$, and
the full expression for the form factors $t_{1},\dots,t_{6}$ are:
\begin{align}
t_{1}= & \int\frac{d^{3}\vec{q}}{(2\pi)^{3}}\frac{\vec{q}^{2}}{mM^{2}\omega}\Bigg\{\bigg[-C_{1}(M_{Z}^{2}+M^{2})-C_{1}'(M_{Z}^{2}-M^{2}+4M\omega)+C_{2}(M_{Z}^{2}-M^{2}-4M\omega)\label{eq:ti}\nonumber \\
 & +C_{2}'(M^{2}+M_{Z}^{2})+C_{3}(2b_{1}M-2M\omega+M_{Z}^{2})+C_{3}'(M_{Z}^{2}-2Md_{1}-2M\omega)\bigg]m\phi_{1}\nonumber \\
 & +\bigg[C_{1}(M^{2}+M_{Z}^{2})+C_{1}'(M_{Z}^{2}-M^{2}+4M\omega)+C_{2}(M_{Z}^{2}-M^{2}-4M\omega)+C_{2}'(M^{2}+M_{Z}^{2})\nonumber \\
 & +C_{3}(2b_{1}M-2M\omega+M_{Z}^{2})+C_{3}'(M_{Z}^{2}-2Md_{1}-2M\omega)\bigg]\omega\phi_{2}\Bigg\}\nonumber \\
t_{2}= & 2t_{6}=2\int\frac{\mathrm{d}^{3}\vec{q}}{(2\pi)^{3}}|\vec{q}|\frac{1}{mM}\biggl[(C_{1}-C_{1}^{'}+C_{2}-C_{2}^{'}+C_{3}-C_{3}^{'})m\phi_{1}+(C_{1}^{'}-C_{1}+C_{2}-C_{2}^{'}+C_{3}-C_{3}^{'})\omega\phi_{2}\biggr],\nonumber 
\\
t_{3}= & \int\frac{d^{3}\vec{q}}{(2\pi)^{3}}\frac{\vec{q}^{2}}{mM^{2}\omega}\Bigg\{\bigg[C_{1}(4M\omega-M^{2}-M_{Z}^{2})-C_{1}'(M_{Z}^{2}-M^{2})+C_{2}(M_{Z}^{2}-M^{2})+C_{2}'(M^{2}+4M\omega+M_{Z}^{2})\nonumber \\
 & +C_{3}(2b_{1}M+2M\omega+M_{Z}^{2})+C_{3}'(M_{Z}^{2}-2d_{1}M+2M\omega)\bigg]m\phi_{1}+\bigg[C_{1}(M_{Z}^{2}-4M\omega+M^{2})+C_{1}'(M_{Z}^{2}-M^{2})\nonumber \\
 & +C_{2}(M_{Z}^{2}-M^{2})+C_{2}'(M^{2}+4M\omega+M_{Z}^{2})+C_{3}(2b_{1}M+2M\omega+M_{Z}^{2})+C_{3}'(M_{Z}^{2}-2d_{1}M+2M\omega)\bigg]\omega\phi_{2}\Bigg\}\nonumber \\
t_{4}= & \int\frac{d^{3}\vec{q}}{(2\pi)^{3}}|\vec{q}|\frac{1}{mM^{2}}\Bigg\{\bigg[C_{1}(2M-4\omega)-C_{2}'(2M+4\omega)+C_{3}(2b_{1}+M+2\omega)+C_{3}'(2d_{1}-M-2\omega)\bigg]m\phi_{1}\nonumber \\
 & +\bigg[C_{1}(4\omega-2M)-C_{2}'(2M+4\omega)+C_{3}(2b_{1}+M+2\omega)+C_{3}'(2d_{1}-M-2\omega)\bigg]\omega\phi_{2}\Bigg\}\nonumber \\
t_{5}= & \int\frac{\mathrm{d}^{3}\vec{q}}{(2\pi)^{3}}\vec{q}^{2}\frac{1}{mM^{2}\omega}\biggl[(C_{1}+C_{1}^{'}-C_{2}-C_{2}^{'}-C_{3}-C_{3}^{'})m\phi_{1}-(C_{1}+C_{1}^{'}+C_{2}+C_{2}^{'}+C_{3}+C_{3}^{'})\omega\phi_{2}\biggr],\nonumber \\
\end{align}
Here, the amplitude is also expanded up to the second order in $|\vec{q}|$.
The behaviors of the coefficients $C_{i}$ and $C_{i}^{'}$ are shown in Fig.~\ref{fig:cchic1},
with $\chi_{c1}(1P)$ taken as an example.

The corresponding decay width is given by: 
\begin{align}
\Gamma_{Z\rightarrow\chi_{c1}\gamma} & =\frac{1}{2^{8}\pi}\biggl(\frac{ee_{Q}gg_{v}^{Q}}{\cos\theta_{W}}\biggr)^{2}\frac{(M_{Z}^{2}-M^{2})}{M_{Z}^{5}}A_{\chi_{c1}}.\label{eq:chic1width}
\end{align}
For brevity, we have introduced $A_{\chi_{c1}}$ in the above expression,
its full expression is given in the Appendix A.

Using the definition in Eq.~\ref{eq:Rn}, We find that the dimensionless
$t_{1,3}$ are of order $R_{2}/M^{3}$. $t_{2,4,6}$ have inverse
mass-energy dimensions and are of order $R_{1}/(MM_{Z}^{2})$. While
$t_{5}$ carries inverse squared mass-energy dimensions and is of
order $R_{2}/(M^{3}M_{Z}^{2})$. Therefore, the two terms in the sum
(Eq.~\ref{eq:Achic1approx}) in the decay width formula are of order
$\dfrac{R_{2}^{2}}{M^{6}M_{Z}}$ and $\dfrac{R_{1}^{2}}{M^{2}M_{Z}}$,
respectively. Notice that higher order terms $\dfrac{R_{2}^{2}M_{Z}}{M^{6}}$
cancels with $M\ll M_{Z}$. Compared with $\chi_{c0}$, the corresponding
decay width is significantly larger.

\subsection{$1^{+-}$ Charmonium}

\begin{figure}[h]
\centering \includegraphics[scale=0.4]{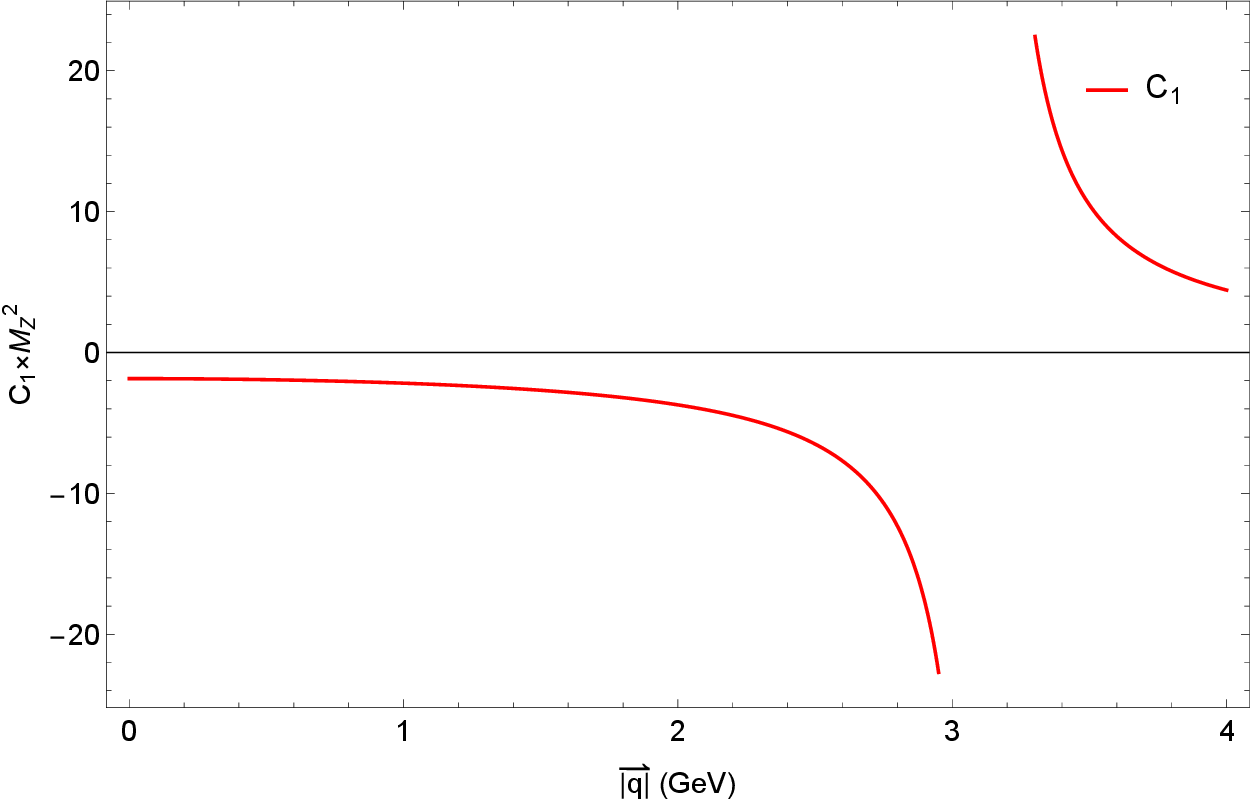}\includegraphics[scale=0.4]{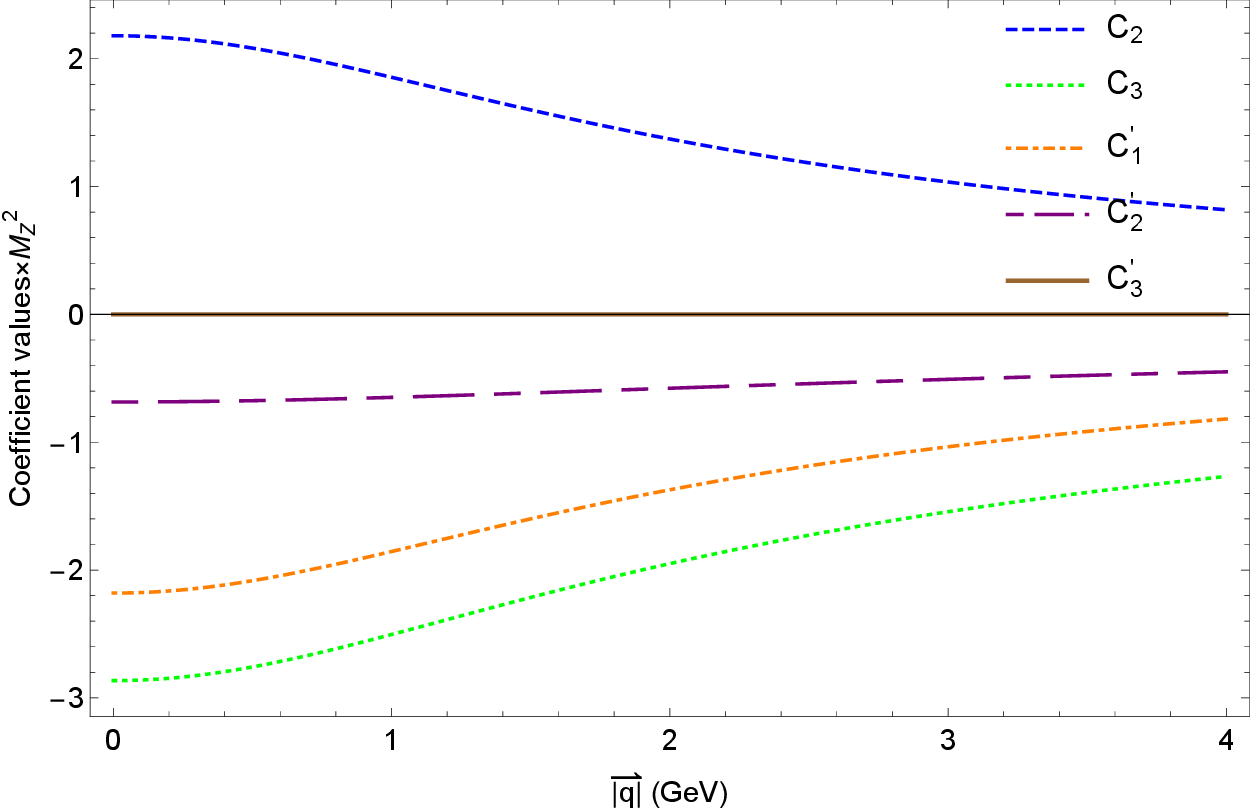}
\caption{ Coefficients $C_{1}$ (left) and $C_{i},\,C_{i}'$ (right) multiplied
by $M_{Z}^{2}$ versus $|\vec{q}|$ in the range $0<|\vec{q}|<4\,\text{GeV}$
for the $h_{c}(1P)$ state. }
\label{fig:chc} 
\end{figure}

Now, we study the processes, $Z\rightarrow h_{c}(mP)+\gamma$ with
$m=1,2$, and calculate their decay widths. To evaluate the decay
widths, we need to introduce the relativistic BS wavefunction; therefore,
we begin with BS wavefunction for the $J^{PC}=1^{+-}$ state. The
general decomposition of the instantaneous BS wavefunction takes the
form \cite{Chang2010}: 
\begin{align}
\psi_{h_{c}}(q_{\perp})= & q_{\perp}\cdot\epsilon_{\perp}^{\lambda}\Bigl[\phi_{1}(q_{\perp})+\phi_{2}(q_{\perp})\frac{\slashed{P}}{M}+\phi_{3}(q_{\perp})\frac{\slashed{P}\slashed{q}_{\perp}}{M^{2}}\Bigr]\gamma_{5}
\end{align}
Here, $P$ and $\epsilon$ are momentum and polarization vector of
$h_{c}$ with mass $M$, respectively. Using the last two equations
of Eq.~\ref{eq:4inde}, we obtain 
\begin{equation}
\phi_{3}(q_{\perp})=-\phi_{2}(q_{\perp})\frac{M}{m}
\end{equation}
As a result, the wavefunction reduces to 
\begin{equation}
\psi_{h_{c}}(q_{\perp})=q_{\perp}\cdot\epsilon_{\perp}^{\lambda}\left[\phi_{1}(q_{\perp})+\phi_{2}(q_{\perp})\left(1+\frac{\slashed{q}_{\perp}}{m}\right)\frac{\slashed{P}}{M}\right]\gamma_{5}.\label{hcwf}
\end{equation}
The corresponding normalization condition is given by: 
\begin{equation}
4\int\frac{\mathrm{d}^{3}\vec{q}}{(2\pi)^{3}}\frac{\phi_{1}(q_{\perp})\phi_{2}(q_{\perp})\omega\vec{q}^{2}}{m}=3P^{0}.
\end{equation}
Using the first two equations of Eq.~\ref{eq:4inde}, we obtain the
coupled integral equations. 
\begin{align}
(M-2\omega)\left[\phi_{1}(q_{\perp})+\phi_{2}(q_{\perp})\frac{\omega}{m}\right] & =\int\frac{\mathrm{d}^{3}\vec{k}}{(2\pi)^{3}}\frac{(\vec{k}\cdot\vec{q})}{\omega m\vec{q}^{2}}\nonumber \\
 & \times\Bigl[(V_{s}+V_{v})\phi_{2}(k_{\perp})(\vec{k}\cdot\vec{q})-(V_{s}-V_{v})\bigl(\phi_{1}(k_{\perp})\omega m+\phi_{2}(k_{\perp})m^{2}\bigr)\Bigr],\\
(M+2\omega)\left[\phi_{1}(q_{\perp})-\phi_{2}(q_{\perp})\frac{\omega}{m}\right] & =\int\frac{\mathrm{d}^{3}\vec{k}}{(2\pi)^{3}}\frac{(\vec{k}\cdot\vec{q})}{\omega m\vec{q}^{2}}\nonumber \\
 & \times\Bigl[(V_{s}+V_{v})\phi_{2}(k_{\perp})(\vec{k}\cdot\vec{q})+(V_{s}-V_{v})\bigl(\phi_{1}(k_{\perp})m\omega-\phi_{2}(k_{\perp})m^{2}\bigr)\Bigr].\label{eq:hc}
\end{align}
One can follow the same steps mentioned in above cases to obtain the
corresponding total BS amplitude via the axial-vector. 
\begin{align}
\mathcal{M}_{Z\rightarrow h_{c}\gamma} & =\frac{\sqrt{3}ee_{Q}gg_{a}^{Q}}{2\cos\theta_{W}}\biggl[I\cdot\epsilon_{1}(y_{1}I\cdot\epsilon_{\gamma}P\cdot\epsilon_{z}+y_{2}\epsilon_{z}\cdot\epsilon_{\gamma})\biggr]
\end{align}
Here, $\epsilon_{1}$ is polarization vector of $h_{c}$, and the
full expression of $y_{1}$ and $y_{2}$ reads as 
\begin{align}
y_{1}= & -4\int\frac{\mathrm{d}^{3}\vec{q}}{(2\pi)^{3}}\frac{\vec{q}^{2}}{M\omega}\biggl[m(C_{1}+C_{1}^{'}-C_{2}-C_{2}^{'}-C_{3}-C_{3}^{'})\phi_{1}+\omega(C_{1}+C_{1}^{'}+C_{2}+C_{2}^{'}+C_{3}+C_{3}^{'})\phi_{2}\biggr]\nonumber \\
y_{2}= & \int\frac{\mathrm{d}^{3}\vec{q}}{(2\pi)^{3}}\frac{|\vec{q}|}{M\omega}\biggl\{\Bigl[C_{1}(M_{Z}^{2}+M^{2}-4M\omega)-C_{1}'(M_{Z}^{2}-M^{2})-C_{2}(M_{Z}^{2}-M^{2})+C_{2}'(M_{Z}^{2}+M^{2}+4M\omega)\nonumber 
\\
 & -C_{3}(M_{Z}^{2}+2M\omega+2Mb_{1})+C_{3}'(M_{Z}^{2}+2M\omega-2Md_{1})\Bigr]m\phi_{1}+\Bigl[C_{1}(M_{Z}^{2}+M^{2}-4M\omega)\nonumber \\
 & -C_{1}'(M_{Z}^{2}-M^{2})+C_{2}(M_{Z}^{2}-M^{2})-C_{2}'(M_{Z}^{2}+M^{2}+4M\omega)+C_{3}(M_{Z}^{2}+2M\omega+2Mb_{1})\nonumber \\
 & -C_{3}'(M_{Z}^{2}+2M\omega-2Md_{1})\Bigr]\omega\phi_{2}\biggr\}
\end{align}
We can calculate the above expression numerically. Similar to the
other cases, $C_{1}$ is discontinuous while the remaining coefficients
continuous in the region of $0<|\vec{q}|<4\,\text{GeV}$. This behavior
is shown in Fig.~\ref{fig:chc}, with $h_{c}(1P)$ taken as an example.

Finally, the corresponding decay width reads as 
\begin{align}
\Gamma_{Z\rightarrow h_{c}\gamma} & =\frac{1}{2^{8}\pi}\biggl(\frac{ee_{Q}gg_{a}^{Q}}{\cos\theta_{W}}\biggr)^{2}\frac{(M_{Z}^{2}-M^{2})}{M_{Z}^{5}}\Bigl[12M_{Z}^{2}y_{2}^{2}+y_{1}^{2}(M_{Z}^{2}-M^{2})^{2}\Bigr]\label{eq:hcwidth}
\end{align}

We perform a similar order analysis as for the $J^{++}$ case. We
find that the dimensionless $y_{1}$ is of order $R_{2}/(MM_{Z}^{2})$,
while $y_{2}$ is dimensionful and of order $R_{1}/M$. Accordingly,
the two terms in the decay width formula (Eq.~\ref{eq:hcwidth})
are of order $\frac{R_{1}^{2}}{M^{2}M_{Z}}$ and $\frac{R_{2}^{2}}{M^{2}M_{Z}^{3}}$,
respectively. This decay width is of the same order as that for the
$\chi_{c1}$ case and also significantly larger than that of $\chi_{c0}$.

\section{Numerical Results}
\label{sec:numres}

By fitting the mass of the ground state, the numerical values of the
parameters are taken from Ref.~\cite{Chang2010}. 
\begin{equation}
\begin{aligned}\mathbbm{e} & =2.7183, & \beta & =0.06\,\text{GeV}, & \lambda & =0.21\,\text{GeV}^{2},\\
m & =1.62\,\text{GeV}, & \Lambda_{\text{QCD}} & =0.27\,\text{GeV} & \theta_{W} & =28.74^{\circ}, & \alpha_{EM} & =1/128.
\end{aligned}
\label{eq:parameters}
\end{equation}
We use the same values for the set of parameters of S-wave and P-wave
states.

By solving the corresponding coupled integral equations for $J^{PC}$
charmonium states as eigenvalue problems, we derive the masses and
wave functions of ground and radially excited states. These masses
are listed in Appendix B. The amplitudes are formulated via form factors
containing 3D integrals over $\mathrm{d}^{3}\vec{q}$, which are computed
numerically. The integrands of these form factors consist of linear
combinations of $C_{i}$ and $C_{i}'$. Among all coefficients, $C_{1}$
stands out as a dominant term with discontinuity and thus requires
dedicated treatment. We therefore accurately locate the discontinuity
points of $C_{1}$ for various charmonium states, which are also summarized
in Appendix B. We adopt semi-analytical regularization to compute
the integrands of the $C_{1}$-related terms in the form factors.
The relevant implementation details are provided in Appendix C.

Finally, we can calculate the decay widths numerically. Subsequently,
the branching fractions can be calculated by adapting $\Gamma/\Gamma_{t}$.
Where $\Gamma_{t}=2.4995$ GeV is the total decay width of the Z boson.

For S-wave states, 
\begin{align}
\mathrm{Br}[Z\to\gamma J/\psi] & =2.98\substack{+0.21+0.067+0.72\\
-0.21-0.027-0.78
}
\times10^{-7},\nonumber \\
\mathrm{Br}[Z\to\gamma\psi'] & =2.18\substack{+0.16+0.037+0.66\\
-0.17-0.036-0.68
}
\times10^{-7},\nonumber \\
\mathrm{Br}[Z\to\gamma\psi(3S)] & =1.37\substack{+0.092+0.031+0.50\\
-0.099-0.030-0.51
}
\times10^{-7},\nonumber \\
\mathrm{Br}[Z\to\gamma\eta_{c}(1S)] & =7.91\substack{+0.42+0.078+0.42\\
-0.44-0.078-0.47
}
\times10^{-9},\nonumber \\
\mathrm{Br}[Z\to\gamma\eta_{c}(2S)] & =5.50\substack{+0.26+0.086+0.35\\
-0.29-0.086-0.38
}
\times10^{-9},\nonumber \\
\mathrm{Br}[Z\to\gamma\eta_{c}(3S)] & =4.01\substack{+0.21+0.092+0.31\\
-0.23-0.090-0.33
}
\times10^{-9}.
\end{align}

For P-wave states, 
\begin{align}
\mathrm{Br}[Z\to\gamma\chi_{c0}(1P)] & =2.12\substack{+0.19+0.023+0.50\\
-0.19-0.023-0.47
}
\times10^{-12},\nonumber \\
\mathrm{Br}[Z\to\gamma\chi_{c0}(2P)] & =2.58\substack{+0.29+0.053+0.69\\
-0.30-0.052-0.64
}
\times10^{-12},\nonumber \\
\mathrm{Br}[Z\to\gamma\chi_{c1}(1P)] & =2.03\substack{+0.17+0.017+0.89\\
-0.18-0.016-0.83
}
\times10^{-9},\nonumber \\
\mathrm{Br}[Z\to\gamma\chi_{c1}(2P)] & =2.73\substack{+0.10+0.035+1.15\\
-0.34-0.034-1.20
}
\times10^{-9},\nonumber \\
\mathrm{Br}[Z\to\gamma h_{c}(1P)] & =2.47\substack{+0.34+0.039+0.90\\
-0.35-0.038-0.83
}
\times10^{-8},\nonumber \\
\mathrm{Br}[Z\to\gamma h_{c}(2P)] & =2.38\substack{+0.42+0.058+0.80\\
-0.42-0.069-0.88
}
\times10^{-8}.
\end{align}

The theoretical uncertainties in the above results are estimated by
varying each input parameter individually within $\pm10\%$ of its
central value. Here, the first uncertainty arises from $\lambda$,
the second from $\beta$, and the third from the quark mass. Individual
uncertainties are combined in quadrature to obtain the total theoretical
uncertainties for these branching ratios. Tables~\ref{tab:brswave}
and~\ref{tab:brpwave} present the branching fractions for S-wave
and P-wave states, respectively, calculated within the BS framework
and other models.

\begin{table}[h]
\centering \resizebox{\linewidth}{!}{ 
\global\long\def\arraystretch{1.5}%
\begin{tabular}{|c|c|c|c|c|c|c|c|}
\hline 
Branching fractions  & BS(our)  & NRQCD(a) \cite{Wang:2023ssg} & LCDA \cite{Luchinsky2017}  & NRQCD(b)  & NRQCD(c) \cite{Sang2023} & CMS \cite{CMS:2024hhg}  & ATLAS \cite{ATLAS:2022rej}\tabularnewline
\hline 
\hline 
$\text{Br}[Z\to\gamma J/\psi]\times10^{7}$  & $2.98_{-0.80}^{+0.75}$  & $1.04_{-0.16}^{+0.18}$  & $0.88$  & 
\vtop{\hbox{$0.996\pm0.186$ \cite{Huang2015}}\hbox{$0.896_{-0.138}^{+0.151}$\cite{Bodwin:2017pzj}
}}  & $0.575_{-0.009}^{+0.008}$  & $<6$  & $<12$\tabularnewline
\hline 
$\text{Br}[Z\to\gamma\psi(2S)]\times10^{7}$  & $2.18_{-0.70}^{+0.68}$  &  &  &  &  & $<13$  & $<24$\tabularnewline
\hline 
$\text{Br}[Z\to\gamma\psi(3S)]\times10^{7}$  & $1.37_{-0.52}^{+0.50}$  &  &  &  &  &  & \tabularnewline
\hline 
$\text{Br}[Z\to\gamma\eta_{c}(1S)]\times10^{9}$  & $7.91_{-0.65}^{+0.60}$  & $13.2_{-5.4}^{+5.6}$  & $9.4$  &  & $9.5_{-0.2}^{+0.2}$  &  & \tabularnewline
\hline 
$\text{Br}[Z\to\gamma\eta_{c}(2S)]\times10^{9}$  & $5.50_{-0.48}^{+0.44}$  &  &  &  &  &  & \tabularnewline
\hline 
$\text{Br}[Z\to\gamma\eta_{c}(3S)]\times10^{9}$  & $4.01_{-0.41}^{+0.38}$  &  &  &  &  &  & \tabularnewline
\hline 
\end{tabular}} \caption{ Branching fractions for $Z\to\gamma\psi(nS)$ and $Z\to\gamma\eta_{c}(nS)$
calculated within the BS model, together with alternative theoretical
predictions and experimental upper limits. The theoretical uncertainties
of our BS results come from the uncertainties of the input parameters
$\lambda$, $\beta$, and $m_{c}$. Here NRQCD(a), NRQCD(b) and NRQCD(c)
stand for calculations from three distinct references, evaluated up
to $O(\alpha_{s}v^{2})$, $O(\alpha_{s}+v^{2})$ and $O(\alpha_{s}^{2})$
including next-to-leading-logarithm resummation, respectively.}
\label{tab:brswave} 
\end{table}

\begin{table}[h]
\centering 
\global\long\def\arraystretch{1.5}%
\begin{tabular}{|c|c|c|c|}
\hline 
Branching fractions  & BS(our)  & NRQCD \cite{Sang2022}  & LCDA \cite{Luchinsky2017} \tabularnewline
\hline 
\hline 
$\text{Br}[Z\to\gamma\chi_{c0}(1P)]\times10^{9}$  & $0.00212_{-0.00051}^{+0.00054}$  & $0.374$  & $0.50$ \tabularnewline
\hline 
$\text{Br}[Z\to\gamma\chi_{c0}(2P)]\times10^{9}$  & $0.00258_{-0.00071}^{+0.00075}$  &  & \tabularnewline
\hline 
$\text{Br}[Z\to\gamma\chi_{c1}(1P)]\times10^{9}$  & $2.03_{-0.85}^{+0.91}$  & $2.38$  & $5.6$ \tabularnewline
\hline 
$\text{Br}[Z\to\gamma\chi_{c1}(2P)]\times10^{9}$  & $2.73_{-1.25}^{+1.16}$  &  & \tabularnewline
\hline 
$\text{Br}[Z\to\gamma h_{c}(1P)]\times10^{9}$  & $24.7_{-9.0}^{+9.6}$  & $3.48$  & $10$ \tabularnewline
\hline 
$\text{Br}[Z\to\gamma h_{c}(2P)]\times10^{9}$  & $23.8_{-9.7}^{+9.0}$  &  & \tabularnewline
\hline 
\end{tabular}
\caption{Branching fractions for the process $Z\to\gamma\chi_{cJ}(nP)(J=0,1)$
and $Z\to\gamma h_{c}(nP)$ calculated in BS model, along with the
previous findings. The theoretical uncertainties associated with our
BS calculations originate from the input parameters $\lambda$, $\beta$,
and $m_{c}$.}
\label{tab:brpwave} 
\end{table}

It is interesting to note that if we adopt the non-relativistic limit of wavefunctions
(by setting $q_{\perp}\sim0$ in Eqs.~\ref{etawf} and \ref{eq:vettf-1}),
the branching fractions are found to be $1.6\times10^{-7}$ and $2.4\times10^{-8}$
for $Z\rightarrow\gamma J/\psi$ and $Z\rightarrow\gamma\eta_{c}$,
respectively. This indicates that relativistic corrections are substantial
and essential for vector charmonium states, while the LO contribution
dominates for pseudoscalar states. Specifically, relativistic effects
enhance the LO prediction by roughly a factor of 2 for $J/\psi$,
and reduce the LO value by around 70\% for $\eta_{c}$. This discrepancy
mainly arises from the different coupling types of the Z boson. The
process $Z\rightarrow\gamma\eta_{c}$ proceeds via vector coupling,
while $Z\rightarrow\gamma J/\psi$ is governed by axial-vector coupling.
In these transitions, the axial-vector coupling is considerably stronger
than the vector coupling.

We now compare our results with predictions from other theoretical
models. Our BS branching fraction for $Z\to\gamma J/\psi$ is roughly
a factor of 3 larger than the NRQCD result computed up to $O(\alpha_{s}v^{2})$,
and about a factor of 3.4 higher than the LCDA prediction. Meanwhile,
our result is still consistent with existing experimental upper limits
\cite{ATLAS:2022rej,CMS:2024hhg}. In contrast, the BS branching fraction
for $Z\to\gamma\eta_{c}$ is roughly 60\% lower than the corresponding
NRQCD prediction. Within the NRQCD framework, the $O(v^{2})$ NLO
relativistic corrections yield negative contributions to both LO and
$\mathcal{O}(\alpha_{s})$ NLO decay widths for $J/\psi$ and $\eta_{c}$
\cite{Wang:2023ssg}. In NRQCD fixed-order computations, relativistic
effects are incorporated perturbatively as $O(v^{2})$ corrections,
whereas the BS formalism is a fully relativistic, Lorentz-covariant
first-principles method that naturally accounts for relativistic effects
to all orders. It would be valuable to compare our BS predictions
with all-order relativistic NRQCD results using the resummation techniques
in \cite{Bodwin:2006dn,Bodwin:2007ga}, which include approximate higher-order
$v^{2}$ matrix elements. Furthermore, it is worthwhile to extend
our BS calculations to NLO in $\alpha_{s}$, since the $O(\alpha_{s}v^{2})$
corrections are found to contribute comparably to, or even more significantly
than, the $O(\alpha_{s}v^{0})$ corrections in NRQCD analyses \cite{Wang:2023ssg}.
However, although higher-order corrections yield significant contributions
to numerous production and decay channels \cite{Zhang:2005cha,Zhang:2006ay,Ma:2010yw,Gong:2012ug,Li:2023tzx,Zhang2021,Zhang:2023mky,Chen2021},
the LO widths dominate for $Z\to\gamma J/\psi,\eta_{c}$. Accordingly,
our LO BS predictions are also expected to be the dominant terms within
the $\alpha_{s}$ expansion.

For the channels of P-wave states, according to the foregoing analysis,
the decay widths of scalar states are suppressed by three powers of the Z boson mass,
or even more strongly. The numerical results in the table clearly
confirm this conclusion. The branching fraction of $\chi_{c0}$ is
smaller than the predictions from NRQCD and LCDA by two orders of
magnitude. In contrast, we can see that the decay width of $\chi_{c1}$
agrees with the NRQCD prediction and is smaller than the LCDA result.
For the $h_{c}$ state, its decay width is larger than the NRQCD value
by a factor of 5 to 10, and roughly 1.5 to 3 times the LCDA result.
Nevertheless, it remains challenging for experimental measurements.

As the principal quantum number increases, the decay widths of S-wave
charmonium channels gradually decrease. In contrast, the decay widths
of P-wave $\chi_{c1,2}$ states increase with radial excitation. These
trends agree with the predictions of conventional potential models.
Meanwhile, we observe a slight reduction for $h_{c}(nP)$, albeit
with relatively larger uncertainties.

\section{Summary }
\label{sec:sum}

We adopted the BS formalism to investigate rare radiative Z-boson
decays into S- and P-wave charmonium, working with LO color-singlet
diagrams. Our core goal was to quantify relativistic corrections to
these processes, so we constructed fully relativistic quarkonium wave
functions and solved the BS equation within the instantaneous approximation.
The BS transition amplitudes take the form of polynomials in the form
factors, with every form factor written as a 3D momentum integral
built from linear combinations of $C_{i}$ and $C_{i}'$. The singular
integral stemming from discontinuous coefficient $C_{1}$ is regularized
semi-analytically with the Sokhotski--Plemelj formula for stable
numerical evaluation.

Our BS calculation yields large branching ratios for $J/\psi$ and
$h_{c}$ final states due to the Z boson’s axial-vector coupling,
while $\chi_{c0}$ decay width is highly suppressed. Relativistic corrections
considerably shift the decay widths and induce noticeable differences
from NRQCD and LCDA results in literatures. For $Z\to\gamma J/\psi$,
our result is enhanced by a factor larger than three compared with
prior predictions, pointing to an underestimation in earlier computations.
This renders our prediction more encouraging for future experimental
searches. These BS-model predictions supply reliable theoretical inputs
for future high-luminosity facilities, such as the ILC, FCC-ee, CEPC,
and Super Z factory.

\section*{Appendix }

\subsection*{A. Expressions for $A_{\psi}$ and $A_{\chi_{c1}}$}

The full expressions for $A_{\psi}$ and $A_{\chi_{c1}}$, which are
defined in the formulas (Eqs.~\ref{eq:psiwidth} and ~\ref{eq:chic1width})
for the decay widths of $1^{--}$ and $1^{++}$ states respectively,
are given below. For convenience, we define $B=M_{Z}^{2}-M^{2}$,
$J_{a}=J_{4}-J_{7}+J_{8}$, $J_{b}=J_{4}-J_{10}=J_{a}-J_{8}$, $J_{c}=J_{2}+J_{5}$,
$J_{d}=J_{2}-J_{6}$, $J_{e}=J_{2}+J_{9}$.

\begin{align}
A_{\psi} & =\Big\{ J_{a}(2J_{a}-J_{b})M^{2}+\Big[3J_{10}^{2}+J_{b}(2J_{4}-J_{a})\Big]M_{Z}^{2}\Big\} M^{2}B^{2}+8\big[J_{11}J_{d}+J_{1}(J_{c}+J_{d}+J_{e})\big]M^{2}M_{Z}^{2}B\nonumber \\
 & +\Big[(J_{1}-J_{c})^{2}+(J_{1}+J_{11}-J_{d})^{2}-4J_{3}J_{a}\Big]M^{2}B^{2}+8\big(J_{c}^{2}+J_{d}^{2}+J_{e}^{2}\big)M^{4}M_{Z}^{2}+2J_{1}^{2}M_{Z}^{2}B^{2}\nonumber \\
 & +J_{3}^{2}\big(B^{2}-8M^{2}M_{Z}^{2}\big)
\end{align}

\begin{align}
A_{\chi_{c1}}= & t_{5}^{2}B^{4}+8t_{3}^{2}M^{2}M_{Z}^{2}+t_{1}^{2}\big(B^{2}+8M^{2}M_{Z}^{2}\big)\nonumber \\
 & +2B^{2}\Big\{\big[t_{1}t_{5}+\big(t_{2}-t_{4}\big)\big(t_{2}-t_{4}-t_{6}\big)\big]M^{2}+\big[t_{1}t_{5}-t_{6}\big(t_{2}-t_{4}-6t_{6}\big)\big]M_{Z}^{2}\Big\}
\end{align}

With $M\ll M_{Z}$ and using the expressions for $J_{i}$ (Eq.~\ref{eq:ji}),
we find 
\begin{equation}
J_{3}\approx-\frac{M_{Z}^{2}}{2}\,J_{4}\label{eq:j3j4}
\end{equation}
Using this relation, we further derive 
\begin{equation}
A_{\psi}\approx\frac{1}{4}M_{Z}^{8}\,J_{4}^{2}\label{eq:Avapprox}
\end{equation}

Similarly, with $M\ll M_{Z}$ and using the expressions for $t_{i}$
(Eq.~\ref{eq:ti}), we find 
\begin{equation}
t_{1}\approx t_{3}\approx-M_{Z}^{2}t_{5}\label{eq:t1t5}
\end{equation}
Using this relation, we further derive 
\begin{equation}
A_{\chi_{c1}}\approx M_{Z}^{6}\bigl[12M^{2}t_{5}^{2}+t_{2}\bigl(t_{4}+2t_{2}\bigr)\bigr].\label{eq:Achic1approx}
\end{equation}

\subsection*{B. Masses Spectra for charmonia and Discontinuity Points of $C_{1}$}

The masses and discontinuity points of $C_{1}$ (denoted by $|\vec{q}|_{0}$)
for different charmonia are presented in Table~\ref{tab:discontinuity_points}.

\begin{table}[htb]
\centering 
\global\long\def\arraystretch{1.2}%
\begin{tabular}{|c|c|c|c|}
\hline 
\textbf{Charmonia (mass in GeV)}  & $\boldsymbol{|\vec{q}|_{0}}$ (GeV)  & \textbf{Charmonia (mass in GeV)}  & $\boldsymbol{|\vec{q}|_{0}}$ (GeV) \tabularnewline
\hline 
\hline 
\multicolumn{4}{|c|}{\textbf{S-wave Charmonium}}\tabularnewline
\hline 
$\eta_{c}(1S)(2.980)$  & 2.4974  & $J/\psi(3.096)$  & 2.6341 \tabularnewline
\hline 
$\eta_{c}(2S)(3.637)$  & 3.249  & $\psi(2S)(3.686)$  & 3.304 \tabularnewline
\hline 
$\eta_{c}(3S)(3.945)$  & 3.588  & $\psi(3S)(4.039)$  & 3.691 \tabularnewline
\hline 
\hline 
\multicolumn{4}{|c|}{\textbf{P-wave Charmonium}}\tabularnewline
\hline 
$\chi_{c0}(1P)(3.414)$  & 3.0005  & $\chi_{c0}(2P)(3.836)$  & 3.4696 \tabularnewline
\hline 
$\chi_{c1}(1P)(3.510)$  & 3.1079  & $\chi_{c1}(2P)(3.871)$  & 3.5080 \tabularnewline
\hline 
$h_{c}(1P)(3.526)$  & 3.1258  & $h_{c}(2P)(3.943)$  & 3.5867 \tabularnewline
\hline 
\end{tabular}\caption{Masses and discontinuity points of $C_{1}$ in $|\vec{q}|$ for S-wave
and P-wave charmonium states.}
\label{tab:discontinuity_points} 
\end{table}

From the table, one can observe that as the mass of the charmonium
increases, the discontinuity point of $C_{1}$ shifts to larger values
of $|\vec{q}|$. This trend can also be derived from the expression
for the discontinuity point. It is given in Appendix C, from which
we see that its value increases with the meson mass for masses smaller
than $M_{Z}$.

\subsection*{C. Regularization of Singular Integrals for $C_{1}$}


The integrands of the form factors associated with $C_{1}$ contain
a singularity originating from the denominator $\bigl(\frac{1}{2}M-\omega-b_{1}\bigr)$.
The simple pole in $C_{1}$ at $|\vec{q}|_{0}$ renders the momentum
integral singular. To handle this singularity, we apply the Sokhotski--Plemelj
formula and decompose $C_{1}$ into a singular pole term and a non-singular
part: 
\begin{align}
C_{1} & =\frac{c_{0}/|\vec{q}|_{0}}{|\vec{q}|-|\vec{q}|_{0}}+\mathcal{C}_{\mathrm{reg}}(|\vec{q}|)\label{eq:C1_decomp}
\end{align}
Here, the coefficient $c_{0}$ in the residue $c_{0}/|\vec{q}|_{0}$
and the pole position $|\vec{q}|_{0}$ are defined as 
\begin{equation}
c_{0}=\omega_{0}^{2}/M_{Z}^{2},\quad|\vec{q}|_{0}=\sqrt{\omega_{0}^{2}-m^{2}},
\end{equation}
where $\omega_{0}=\frac{MM_{Z}^{2}}{M^{2}+M_{Z}^{2}}$. Substituting
this decomposition into the momentum integral, we split the original
singular integral into a singular Cauchy principal value part and
a regular non-singular part that can be evaluated numerically without
divergence issues: 
\begin{align}
\int\frac{\mathrm{d}^{3}\vec{q}}{(2\pi)^{3}}C_{1}f(q_{\perp}) & =\int\frac{\mathrm{d}|\vec{q}|}{2\pi^{2}}\frac{c_{0}|\vec{q}|_{0}}{|\vec{q}|-|\vec{q}|_{0}}f(|\vec{q}|_{0})+\int\frac{\mathrm{d}|\vec{q}|}{2\pi^{2}}\Big[\vec{q}^{2}C_{1}f(|\vec{q}|)-\frac{c_{0}|\vec{q}|_{0}}{|\vec{q}|-|\vec{q}|_{0}}f(|\vec{q}|_{0})\Big].
\end{align}
For the singular term, the angular integration is trivial, leaving
only the radial integral, which is interpreted as the Cauchy principal
value to obtain a physically finite outcome. Here $u_{q}$ denotes
the upper cutoff of the physical momentum space, and the singular
radial integral yields a logarithmic expression: 
\begin{align}
\int\frac{\mathrm{d}|\vec{q}|}{2\pi^{2}}\frac{c_{0}|\vec{q}|_{0}}{|\vec{q}|-|\vec{q}|_{0}}f(|\vec{q}|_{0})=\frac{c_{0}|\vec{q}|_{0}f(|\vec{q}|_{0})}{2\pi^{2}}\mathrm{P.V.}\int_{0}^{u_{q}}\frac{\mathrm{d}|\vec{q}|}{|\vec{q}|-|\vec{q}|_{0}}=\frac{c_{0}|\vec{q}|_{0}f(|\vec{q}|_{0})}{2\pi^{2}}\ln\frac{u_{q}-|\vec{q}|_{0}}{|\vec{q}|_{0}}.
\end{align}
Remarkably, the dependence on the upper momentum cutoff $u_{q}$ cancels
out entirely between the above expression and the non-singular integrand.
Consequently, the final result is cutoff-independent for sufficiently
large $u_{q}$.


\section*{Acknowledgements}

This work was supported by the National Natural Science Foundation
of China (Grants No. 11705078, No. 12575087).

\bibliographystyle{unsrt}
\addcontentsline{toc}{section}{\refname}

\end{document}